\documentclass[12pt]{article}
\renewcommand{\thefootnote}{\fnsymbol{footnote}} 
 
\def\g{\Gamma} 
\def\gg{{\rm I}\!\g} 
\newcommand{\fp}{\Phi\Pi} 
\newcommand{\alp}{\alpha}

\newcommand{\nabh}{\hat{\nabla}} 
 
\newcommand{\gh}{\hat{\g}}

\def\cf{{\cal F}} 
\def\xic#1{\xi_{_{#1}}}

\begin{document} 
\renewcommand {\theequation}{\arabic{section}.\arabic{equation}} 
\begin{flushright} 
NYU-TH/00/09/09\\ 
CERN-TH/2001-002\\ 
DESY 01--008\\ 
January 2001\\ 
\end{flushright} 
 
\vskip 0.5cm 
\begin{center} 
  \centerline{\LARGE\bf  The Algebraic Method} 
  \vskip 2.0cm 
{ 
  {\large\bf P.A. Grassi}$^{a}$, 
  {\large\bf T. Hurth}$^{b}$, and {\large\bf M. Steinhauser}$^{c}$ 
  \\[.9cm] 
  {\it (a) New York University, Department of Physics,\\ 
   4 Washington Place, New York, NY 10003, USA} 
  \\[.5em] 
  {\it (b) CERN, Theory Division,\\ 
   CH-1211 Geneva 23, Switzerland.} 
  \\[.5em] 
  {\it (c) II. Institute for Theoretical Physics, 
   University of Hamburg,\\ 
    Luruper Chaussee 149, 22761 Hamburg, Germany.}\\ 
} 
 
\vskip 1.5cm 
  \begin{minipage}{14.8cm} 
    {\small 
      \begin{center} 
        {\bf Abstract} 
      \end{center} 
      \vskip 0,2cm 
Combining the effect of an intermediate renormalization 
prescription (zero momentum subtraction) and the background 
field method (BFM), we show that the algebraic renormalization 
procedure needed for the computation of radiative corrections 
within non-invariant regularization schemes is 
drastically simplified. The present technique is suitable for 
gauge models and, here, is applied to the Standard Model. The 
use of the BFM allows a powerful organization of the counterterms and 
avoids complicated Slavnov-Taylor identities. 
Furthermore, the 
Becchi-Rouet-Stora-Tyutin (BRST) variation of background 
fields plays a special role in disentangling Ward-Takahashi 
identities (WTI) and 
Slavnov-Taylor identities (STI). Finally, the strategy 
to be applied to physical processes is exemplified for the 
process $b\rightarrow s\gamma$. } 
  \end{minipage} 
 
\end{center} 
\thispagestyle{empty} 
\newpage 
\setcounter{page}{1} 
\renewcommand{\thefootnote}{\arabic{footnote}} 
\setcounter{footnote}{0} 
 

\setcounter{equation}{0} 
\section{Introduction} 
\label{sec:introduction} 
 
The method of algebraic renormalization has been intensively used 
as a tool for proving renormalizability of various 
models (see, e.g.,~\cite{libro}). However, its full value has not yet 
been widely appreciated by the practitioners.  Indeed, the theoretical 
understanding of algebraic renormalization does not lead automatically 
to a practical advice for higher-loop calculations. 
We have to recall that the quantization of gauge models in covariant 
gauges is necessarily characterized by complicate STIs
among the Green functions. Already for simple models, the 
algebraic renormalization --- which essentially provides a  
technique to restore the symmetries broken by non-invariant  
regularization schemes --- appears particularly 
difficult~\cite{martin}. In the case of the Standard Model (SM)   
or its supersymmetric extensions, the number of 
fields, of couplings and of independent renormalization constants forbid 
the naive application of the method~\cite{STII,krau_ew,Maggiore,hollik}. 
 
In a recent paper~\cite{amt_1}, the algebraic renormalization 
has been considered with regard to practical 
applications. A procedure has 
been suggested and worked out which allows an efficient determination 
of the breaking terms and of the corresponding counterterms. 
Actually the computation can be reduced to the 
evaluation of universal, i.e. regularization-scheme-independent, 
counterterms. 
 
In this paper, we present an improved strategy to simplify the 
renormalization program within a non-invariant regularization 
scheme by using  the advantages of the  
BFM~\cite{bkg,msbkg,grassi}. 
The latter simplifies consistently the analysis and provides an 
effective procedure for computing all necessary counterterms. 
 
The difference between the conventional approach without the BFM  
and the new method presented here is 
essentially due to the fact that the number of independent 
breaking terms is highly reduced. 
This is achieved by implementing the 
WTIs also for the background fields.  
In a first step the corresponding counterterms are computed, which is 
relatively simple due to the linearity of the WTIs.
The counterterms are in turn incorporated into the  
STIs where, as a consequence, 
only a few background gauge invariant counterterms are needed 
to restore the non-linear equations. 
Thus we completely avoid the use of 
the conventional STIs, obtained by differentiating the functional STI with 
respect to the quantum fields and at least one ghost field. 
However, we cannot totally skip the STIs obtained with at least one 
anti-field. 
Next to the presence of the BRST variation of the background fields 
the other main new aspect of our method
is the effective solution of the remaining 
STIs by exploiting the background gauge invariance. 
 
For many models, it is quite easy to find a regularization technique
which preserves background gauge invariance at one- or two-loop order
due to
the linearity of the WTIs. In such cases, the only missing counterterms are
those needed to
restore the STIs. Owing the background gauge invariance, the number of
independent
non-invariant counterterms are reduced and, according to our procedure, the
program can be further simplified. In addition, exploiting completely the BFM,
only one-loop non-invariant counterterms are really necessary to perform
the computation of physical amplitudes.

The practical use of a such an optimized algebraic scheme is 
at least two-fold: the impressive experimental precision mainly reached 
at the electron-positron colliders LEP and SLC and at the proton 
anti-proton collider TEVATRON has made it 
mandatory to evaluate higher-order quantum 
correction where the algebraic inconsistencies within the naive 
dimensional regularization scheme  are unavoidable. It also 
seems desirable to have a powerful alternative for 
cross-checks. Moreover, 
using the dimensional scheme in supersymmetric theories 
one needs a practical procedure to restore the Ward 
identities of supersymmetry in the final step 
of the renormalization procedure. 
 
In \cite{amt_2}, we applied the method of Ref.~\cite{amt_1}
to the three-gauge-boson vertices involving two $W$ bosons 
and a photon or $Z$ boson.  
These contributions 
constitute a building block to the important $W$ 
pair production process $e^+e^-\to W^+W^-$, which plays a crucial role at 
LEP2. 
Also the two-loop electroweak muon decay
amplitude was discussed within this approach~\cite{AT}.
The application of these techniques to supersymmetric 
examples will be presented in a forthcoming paper.
 
The paper is organized as follows: In Section~\ref{sec:sti} 
we recall the quantization 
within the BFM and introduce our conventions. 
In Section~\ref{sub:QAP} we discuss the conventional algebraic method. 
In Section~\ref{sub:remove} our method for the computation of
the breaking terms $\Delta^{(n)}$ to all orders is presented. 
This technique, used in conjunction 
with  the BFM, simplifies the task of the complete analysis. 
It  is based on intermediate renormalization prescriptions. 
In Section~\ref{sub:strategy}, we describe the structure of the 
complete set of identities necessary to renormalize the SM and how 
they should be organized in order to provide the most efficient 
procedure in practical computations. 
Finally, in Section~\ref{sub:application}, we apply 
the previous analysis to the physical process $b\to s\gamma$
and present our conclusions in Section~\ref{sec:con}.
In Appendix~\ref{app:linear} 
we discuss the linearized Slavnov-Taylor operator and the couplings
not presented in~\cite{amt_1}. 
Besides the WTIs and STIs there exist also other functional identities
which are discussed in Appendix~\ref{app:ghost_equations}. There, also the
reduced functional is briefly introduced.
All possible background gauge invariant counterterms, which in
our method could occur to fix the STIs, are listed in Appendix~\ref{app:cou}.

 
\setcounter{equation}{0} 
\section{Quantization within the Background Field Gauge} 
\label{sec:sti} 
 
For the reader's convenience and in order to establish our 
conventions, we briefly illustrate the quantization procedure for 
the SM in the background field gauge. We recall the BRST symmetry 
(extended to background fields)~\cite{brs,bkg,grassi,krau_ew} and the 
corresponding STIs, the gauge fixing and, finally, the WTIs for 
the background gauge invariance. 
Auxiliary material and supplementary 
constraints such as the Abelian Anti-ghost equation~\cite{grassi} 
are discussed in Appendix~\ref{app:ghost_equations}. For more details
we refer to~\cite{amt_1}. 
 
A generic field is denoted by $\phi$, while  $\Phi_i$ stands for 
scalar matter fields, i.e. Goldstone ($G^\pm$, $G^0$) and Higgs 
($H$) bosons. A generic gauge boson field is denoted by 
$V_{\mu}^B$ and the ghost and the anti-ghost fields by $c^B$ and 
$\bar{c}^B$, respectively. The index $B$ collectively denotes the 
adjoint representation for the group $SU(3)\times SU(2)\times 
U(1)$. 
The symbols $V^a_\mu$ and $c^a$ are used to denote gluon 
fields and the corresponding ghosts in the adjoint representation 
of the Lie algebra $su(3)$.  The background fields are marked with 
a hat in order to distinguish them from their quantum 
counterparts. $Q_i$ denotes the electric charge of a quark $q_i$. 
The Greek indices belong to the adjoint representation for $su(2)$ 
and Latin indices of the beginning of the alphabet $a,b,c,\dots$ 
run over the adjoint representation of the Lie algebra $su(3)$. 
Latin indices from the middle of the alphabet $i,j,k,\dots$ denote 
the representation of scalars which is taken to be real 
($t^A_{ij} = -t^A_{ji}$). The generators 
$t^A$ satisfy the commutation relations $[t^A,t^B] = f^{ABC} 
t^C$ where $f^{ABC}$ denotes the structure constant of the gauge 
group. We use the symbol $\psi_I$ for fermions and 
the generators in the fermion representation are identified by 
${T}^{A}_{IJ}$ where the indices $I,J,K,\ldots$ collect the spin, 
isospin, colour and chirality degrees of freedom. 
 
Let us also introduce three different types of effective actions, which 
will be used in the following. The Green functions $\Gamma$ are 
regularized and subtracted by means of any chosen scheme. 
The Green functions $\hat{\Gamma}$ are 
subtracted using Taylor expansion (see Section~\ref{sub:remove}). 
This implies that all sub-divergences are already renormalized properly.
Finally, $\gg$ denotes the renormalized symmetric Green functions, 
which satisfy the relevant WTIs and STIs.  A complete explanation of 
the conventions, quantum numbers and symmetry transformations of the 
fields is provided in Ref.~\cite{amt_1}. 
 
The quantization of the theory can only be achieved by introducing a suitable 
gauge fixing ${\cal L}_{GF}$ and the corresponding Faddeev-Popov terms 
${\cal L}_{\Phi\Pi}$. Then the classical action reads  
\begin{eqnarray}   
  \label{gree.5} 
  \gg_0 &=& \g_0 \,\,=\,\, \int {\rm d}^4x \left( 
  {\cal L}_{INV} + \sum_i \phi^*_i \, s \phi^i + {\cal L}_{GF} +{\cal 
  L}_{\Phi\Pi} \right) 
  \,. 
\end{eqnarray} 
Both ${\cal L}_{GF}$ and ${\cal L}_{\fp}$ break the local gauge invariance, 
leaving the theory invariant under the BRST~\cite{brs} transformations. 
The BRST symmetry is crucial for proving the unitarity of the S-matrix and 
the gauge independence\footnote{The BRST symmetry
  extended to background fields 
  is essential for quantizing a gauge model in the
  background gauge. Within that framework the unitarity of
  the model, the gauge
  parameter independence and the independence of the S-matrix from 
  the background fields can be proven~\cite{AGS}.}
of physical observables. Therefore it must 
be implemented to all orders. 
Owing to the non-linearity of the 
BRST transformations~\cite{brs}, the renormalization of some composite 
operators (namely $s \phi_i$ where $\phi_i$ is a 
generic field of the SM and $s$ is the BRST generator) is necessary. 
This is usually done by adding the composite operators 
$s \phi_i$ coupled to BRST-invariant external sources $\phi^*_i$ (anti-fields) 
to the classical action, namely we add the sum $\sum_i \phi^*_i \, s \phi^i$, 
which is restricted to those fields undergoing 
a non-linear transformation~\cite{zinn}. 
 
To renormalize properly the SM in the background gauge, one 
needs to implement the equations of motion for the background fields at the 
quantum level. 
The most efficient way to this end is to extend the BRST symmetry to the 
background fields
\begin{eqnarray}
  \label{new_1}
  \begin{array}{llll}
    s \hat{W}^{3}_{\mu} = \Omega^{3}_{\mu}\,,
    &
    s \Omega^{3}_{\mu} = 0\,, 
    &
    s \hat{G}^{0}  = \Omega^{0}\,,
    &
    s \Omega^{0} = 0\,,
    \\
    s \hat{W}^{\pm}_{\mu} = \Omega^{\pm}_{\mu}\,,
    &
    s \Omega^{\pm}_{\mu} = 0\,,
    &
    s \hat{G}^{\pm}  = \Omega^{\pm}\,,
    &
    s \Omega^{\pm} = 0\,,
    \\
    s\hat{G}_{\mu}^{a}  = \Omega^{a}_{\mu}\,,
    &
    s \Omega^{a}_{\mu} = 0\,,
    &
    s \hat{H}  = \Omega^{H}\,,
    &
    s \Omega^{H} = 0\,,
  \end{array}
\end{eqnarray}
where $ \Omega^{\pm}_\mu,  \Omega^{3}_\mu$ and $\Omega ^a_\mu$ are
(classical) vector fields with  
the same quantum numbers as the gauge bosons $W,Z$ and $G^a_\mu$, but
ghost charge $+1$ (like a ghost field).  
$\Omega^{\pm},  \Omega^{0}$ and $\Omega^H$ are scalar fields with
ghost number $+1$.

Finally, 
the BRST symmetry 
extended to the background fields (cf., e.g.,~\cite{amt_1}) 
is implemented at the 
quantum level by establishing the corresponding  STI in the functional form 
\begin{eqnarray}\label{ST} 
  {\cal S}(\gg)[\phi] & = &\int\,\,{\rm d}^4x  \Bigg\{ \Bigg[ 
  \left( s_W \partial_\mu c_Z + c_W \partial_\mu c_A \right) 
  \left( s_W \frac{\delta\gg}{\delta Z_\mu} +  c_W \frac{\delta\gg}{\delta 
      A_\mu} \right) 
  \nonumber\\&&\mbox{} 
+ \frac{\delta \gg}{\delta W^{*,3}_\mu} 
  \left( c_W \frac{\delta\gg}{\delta Z_\mu} -  s_W \frac{\delta\gg}{\delta 
      A_\mu} \right) 
+\frac{\delta \gg}{\delta W^{*,\pm}_\mu}  \frac{\delta\gg}{\delta W^\mp_\mu} 
+ \frac{\delta \gg}{\delta V^{*,a}_\mu}  \frac{\delta\gg}{\delta V^a_\mu} 
+ \frac{\delta \gg}{\delta c^{*,\pm}}  \frac{\delta\gg}{\delta c^\mp} 
  \nonumber\\&&\mbox{} 
+ \frac{\delta \gg}{\delta c^{*,3}} 
\left(c_W \frac{\delta\gg}{\delta c_Z} -  s_W \frac{\delta\gg}{\delta c_A} 
\right) 
+\frac{\delta \gg}{\delta c^{*,a}}  \frac{\delta\gg}{\delta c^a} 
+\frac{\delta \gg}{\delta G^{*,\pm}}  \frac{\delta\gg}{\delta G^\mp} 
+ \frac{\delta \gg}{\delta G^{*,0}}  \frac{\delta\gg}{\delta G^0} 
  \nonumber\\&&\mbox{} 
+ \frac{\delta \gg}{\delta H^*}  \frac{\delta\gg}{\delta H} 
+ \sum_{I = L,Q,u,d,e} \left( 
    \frac{\delta \gg}{\delta \bar{\psi}^{*I}} \frac{\delta\gg}{\delta 
      \psi^{I}} + 
    \frac{\delta \gg}{\delta \psi^{*I}} \frac{\delta\gg}{\delta 
      \bar{\psi}^{I}}
  \right) + \sum_{\alp=A,Z,\pm,a} b_{\alp}  \frac{\delta\gg}{\delta 
    \bar{c}^{\alp}} 
  \Bigg] 
 \nonumber\\&&\mbox{} 
+ \Omega^{3}_\mu 
  \left[ c_W  \left(   
  \frac{\delta\gg}{\delta \hat{Z}_\mu} -   
  \frac{\delta\gg}{\delta Z_\mu}\right) 
 -  s_W \left(  
  \frac{\delta\gg}{\delta \hat{A}_\mu}      -   
  \frac{\delta\gg}{\delta A_\mu} \right) \right] 
\nonumber\\&&\mbox{} 
+\Omega^{\pm}_\mu \left(  \frac{\delta\gg}{\delta \hat{W}^\pm_\mu} - 
\frac{\delta\gg}{\delta W^\pm_\mu}\right) 
+\Omega ^a_\mu  \left(  \frac{\delta\gg}{\delta \hat{V}^a_\mu} - 
\frac{\delta\gg}{\delta V^a_\mu}\right) 
\nonumber\\&&\mbox{} 
+\Omega^{\pm} \left(  
  \frac{\delta\gg}{\delta \hat{G}^\pm} - \frac{\delta\gg}{\delta G^\pm}\right) 
+\Omega^0  \left(  
  \frac{\delta\gg}{\delta \hat{G}^0} -  \frac{\delta\gg}{\delta {G}^0}\right) 
+\Omega^H  \left(  
  \frac{\delta\gg}{\delta \hat{H}} - \frac{\delta\gg}{\delta H}\right)\Bigg\} 
 \nonumber\\ 
  &=& 0 
  \,, 
\end{eqnarray} 
where the notation $A^\pm B^\mp = A^+ B^- + A^- B^+$ has been used. 
Here $s_W$ and $c_W$ denote the sine and cosine of the Weinberg angle 
$\theta_W$, and $b_{\alp}$ are the so-called Nakanishi-Lautrup 
multipliers. 
 
The first sum in the fourth line of Eq.~(\ref{ST}) includes the left-handed 
doublets and the right-handed singlets.  For the BRST source fields no 
Weinberg rotation has been introduced.  We stress that this formula 
represents the complete non-linear STI to all orders. The first two and 
the last term in the first square brackets 
correspond to the linear BRST variation of the $U(1)$ 
abelian gauge field and the BRST transformations of the anti-ghost 
fields. 
 
In order to specify the gauge-fixing,  we introduce the equation of motion 
for the $b$ fields corresponding to the various gauge fields in the SM 
\begin{eqnarray}
  \label{eq:gau_fix} 
  \frac{\delta \gg}{\delta b_C} &=& 
  {\cal F}_C({V}, {\Phi}_i) + \xi_C b_C 
  \,, 
\end{eqnarray} 
where ${\cal F}_C$ are the gauge fixing 
functions. $\xi_C$ ($C= A, Z, \pm, a$) are the corresponding gauge 
parameters.  In the case of the background gauge fixing 
the functions ${\cal F}_C$ are explicitly given in formula 
(A.2) of \cite{amt_1}. 
 
For each generator of the gauge group 
$SU(3) \times SU(2) \times U(1)$, we consider the corresponding 
local infinitesimal parameters. They are denoted by 
$\lambda_A(x), \lambda_Z(x)$ and $\lambda_\pm(x)$ 
for the electroweak part and by $\lambda_a(x)$ for the QCD sector. 
The functional WTI for the effective action $\gg$ thus reads
\begin{eqnarray} 
  \label{WT} 
  {\cal W}_{(\lambda)}(\gg) &=& \sum_{\phi} \int {\rm d}^4x 
  \left(\delta_{\lambda(x)}\phi\right) 
  \frac{\delta 
    \gg}{\delta \phi(x)} = \sum_{B = A,Z,\pm, a} 
  \int {\rm d}^4x \lambda_B(x) {\cal W}_{B}(x) (\gg) \,\,=\,\, 0 
\,, 
\end{eqnarray} 
where the variations $\delta_{\lambda}\phi(x)$ 
are explicitly given in the Appendix of \cite{amt_1}. 
The sum runs over all possible fields and anti-fields. 
${\cal W}_{\alpha}(x)$ is 
called Ward-Takahashi operator. It acts on the 
functional $\gg[\phi]$. 
 
The principal difference between the STIs~(\ref{ST}) and the WTIs~(\ref{WT}) 
for the background gauge invariance is due to the linearity of the latter. 
This means that the WTIs are linear in the functional $\gg$ and 
they therefore relate Green functions of the same orders while for the 
STIs there is an interplay between higher- and lower-order radiative 
corrections. 
 
In Appendix~\ref{app:ghost_equations},  
we consider the supplementary equations needed in order to 
complete the algebraic structure of the functional equations.  
Although these constraints are
not independent from~(\ref{ST}),~(\ref{eq:gau_fix})  and~(\ref{WT}) 
they turn out to be useful in order to  
reduce the complexity of the algebraic method. 

As a final remark of this section we want to mention that 
in the forthcoming part of this paper
instead of Eq.~(\ref{gree.5}) we will use the reduced functional
which differs only slightly from $\Gamma_0$
in the gauge-fixing part as 
is explained explicitly in Appendix~\ref{app:ghost_equations}
(see Eqs.~(\ref{eq:red_func}) and~(\ref{eq:an_red_func})).
This has the consequence that some of the formulae are simpler.
In particular, in Eq.~(\ref{ST}) the last term in the fourth line is
absent. 
As this is a minor technical point and confusion is excluded, we denote
also the reduced functional by $\Gamma_0$.


\setcounter{equation}{0} 
\def\ton{^{(n)}} 
\section{Renormalization} 
 
\subsection{The algebraic method} 
\label{sub:QAP} 
 
The Quantum Action Principle (QAP)~\cite{QAP}
is the fundamental theorem of renormalization theory. It 
guarantees the locality of the counterterms and the polynomial 
character of the renormalization procedure. The QAP also implies that 
all breaking terms of the STIs and WTIs are local and that they can be 
fully characterized in terms of classical fields, their quantum 
numbers and symmetry properties. 
 
In the case of STIs, the QAP implies that, if  the Green functions 
$\gg^{(n-1)}$ satisfy all the symmetry 
constraints at lower orders, the (subtracted)  $n^{\rm th}$-order Green 
functions $\g^{(n)}$ fulfill them up to 
local insertions $\Delta^{(n)}_{S}$ in the one-particle-irreducible
(1PI) Green functions: 
\begin{eqnarray}
  \label{qap_1a} 
  \left[{\cal S}( \g )\right]^{(n)} &=& 
  \Delta^{(n)}_{S} 
  \,. 
\end{eqnarray} 
Here $\Delta^{(n)}_{S}$ is an integrated, Lorentz-invariant polynomial 
of the fields, of the anti-fields and of their derivatives, 
$\Delta^{(n)}_{S} = \sum_i \gamma\ton_{S,i} 
 \int {\rm d}^4x \,\Delta_{S,i}(x)$,
with ultra-violet (UV) degree $\leq 4$ and infra-red 
(IR) degree $\geq 3$ (assuming four space-time dimensions). In the 
same way, the WTIs are spoiled by breaking terms of the form 
\begin{eqnarray}
  \label{qap_1b} 
  {\cal W}_{(\lambda)}\left( \g^{(n)} \right) &=& 
  \Delta^{(n)}_{W}(\lambda) 
  \,, 
\end{eqnarray} 
where  $\Delta\ton_{W}(\lambda)= 
 \sum_i \gamma\ton_{W,i} \int {\rm d}^4x \,\Delta_{W,i}(x,\lambda)$ 
is again an integrated, Lorentz-invariant polynomial, 
depending linearly on the infinitesimal parameter $\lambda$, 
with UV degree $\leq 4$ and IR degree $\geq 3$. 

Although Eqs.~(\ref{qap_1a}) and~(\ref{qap_1b}) apply to any 
renormalization scheme, the coefficients 
$\gamma\ton_{S,i}$ and $\gamma\ton_{W,i}$ 
of the various $\Delta\ton$'s depend on the particular scheme adopted and 
on the order of the computation. In fact, the definitions of 
$\Delta\ton_{S}$ and $\Delta\ton_{W}(\lambda)$ rely on specific 
conventions for composite operators. Thus a renormalization 
description for the composite operators is necessary. Usually one uses 
the concept of Normal Product Operators introduced by 
Zimmermann~\cite{zimm} 
or the conventional counterterm technique, 
which is preferable from the practical point of view. 

Once the breaking terms $\Delta\ton_{S}$ and $\Delta\ton_W(\lambda)$ 
are given, the main objective of the algebraic 
method~\cite{brs,libro} can be discussed. This essentially entails 
a prescription to restore the identities by suitable local 
counterterms\footnote{See Appendix~\ref{app:cou} for the
  notation.}, 
$\g^{CT,(n)} = \sum_i \xi\ton_i \int {\rm d}^4x \, 
{\cal L}^{CT}_i(x)$,
such that one has at $n^{\rm th}$ order: 
\begin{eqnarray}
  \label{qap_new_1} 
  \Big[{\cal S}(\gg)\Big]^{(n)} 
  &\equiv& 
  {\cal S}_0(\g^{(n)}) + \sum^{n-1}_{j=1} 
  \left(\gg^{(j)},\gg^{(n-j)}\right) - \ {\cal S}_0(\g^{CT,(n)}) 
  \,\,=\,\,  0
  \,, \nonumber \\ 
  {\cal W}_{(\lambda)}\left(\gg^{(n)}\right) 
  &\equiv& 
  {\cal W}_{(\lambda)}\left(\g^{(n)}\right) 
  - {\cal W}_{(\lambda)}\left(\g^{CT,(n)}\right) 
  \,\,=\,\,  0
  \,, 
\end{eqnarray} 
where the decomposition given in Eqs.~(\ref{line_ST})
and~(\ref{bracket})  has been used. 
Notice that, since the 
Green functions $\gg^{(j)}$ with $j < n$ are already fixed, 
only $\g^{(n)}$ has to be adjusted by the local counterterms $\g^{CT,(n)}$. 

In practice the problem amounts to solving the algebraic equations 
\begin{eqnarray}
  \label{qap_new_2} 
  {\cal S}_0\left(\g^{CT,(n)}\right) \,\,=\,\,  \Delta\ton_{S}
  \,, 
  \hspace{1.5cm} 
  {\cal W}_{(\lambda)}\left(\g^{CT,(n)}\right) \,\,=\,\,  
  \Delta\ton_{W}(\lambda)
  \,, 
\end{eqnarray} 
where ${\cal S}_0$ is defined below Eq.~(\ref{bracket}). 
The solution fixes a subset of the coefficients 
$\xi_i\ton$ of the counterterms 
in terms of $\gamma\ton_{S,i}$ and $\gamma_{W,i}\ton$. 
These equations turn out to be solvable in the absence of 
anomalies, where only the consistency conditions 
have to be used. Moreover, because of a non-trivial kernel of the 
operators ${\cal S}_0$ and ${\cal W}_{(\lambda)}$ 
(i.e. the space of invariant counterterms), 
one is allowed to impose renormalization conditions by 
tuning the free parameters of the model (namely the remaining coefficients 
$\xi\ton_i$ of $\g^{CT,(n)}$) 
\begin{eqnarray} 
  \label{qap_new_3} 
  {\cal N}_{\cal I} \left( \gg\ton \right)
  &=&   
  {\cal N}_{\cal I}\left( \g\ton + \g^{CT,(n)}\right) 
  \,\,=\,\, 0
  \,, 
\end{eqnarray} 
where the index ${\cal I}$ runs over all independent parameters of 
the SM and ${\cal N}_{\cal I}$ denotes the normalization 
condition operators. 
 
The existence and uniqueness of $\g^{CT,(n)}$ as a solution of the system
(\ref{qap_new_1})--(\ref{qap_new_3}) has been proven in~\cite{brs,coho,glenn}
for gauge theories, in~\cite{ps,hollik} for supersymmetric models, and
in~\cite{STII,krau_ew,grassi} for non semi-simple models coupled
to fermion and scalar fields. Indeed, the main purpose of the algebraic
renormalization is
to demonstrate the existence of a unique solution (up to normalization
conditions) to the algebraic problem  (\ref{qap_new_1})--(\ref{qap_new_2})
in the absence of anomalies.
Unfortunately, this does not necessarily provide a practical
technique  to compute the breaking terms
$\Delta\ton$ and to determine the corresponding counterterms $\g^{CT,(n)}$.

 
\subsection{Simplifying the breaking terms} 
\label{sub:remove} 
 
As already mentioned above, regardless of 
which regularization scheme is used, 
the calculation of the $\Delta\ton$'s in~(\ref{qap_new_2}) is quite 
tedious and gets even more complicated at higher orders 
(see, e.g.,~\cite{martin} for a complete one-loop computation). 
In general, one has to calculate all Green functions that 
occur in the complete set of STIs or WTIs. 
Inserting them in the chosen identities then determines $\Delta\ton$. 
The additional computations necessary in the conventional algebraic method 
can be slightly reduced: instead of calculating all Green functions 
that occur in the full set of STIs, one can 
compute only those at special points, namely for zero 
momentum, for on-shell momentum or for large external momenta. 
As a consequence, the breaking terms, $\Delta\ton$, are simply 
related to those Green functions that are evaluated at 
these special points. 
Clearly, if on-shell renormalization conditions are used in the 
calculation, the on-shell method could be superior to the 
zero-momentum subtraction. At large momentum, one 
takes advantage of Weinberg's theorem~\cite{wein}. 
In~\cite{fg}, based on the zero-momentum subtraction, 
a procedure has been formulated to discuss the 
complete renormalization of the abelian Higgs model. Here, we 
adopt the same procedure in the more general context of the SM. 
 
Owing  to locality and to bounded mass dimension 
(as fixed by power counting and the QAP) of each single monomial 
$\Delta_{S,i}(x)$ and $\Delta_{W,j}(x,\lambda)$ of 
$\Delta\ton_S$ and $\Delta\ton_{(\lambda)}$, respectively, 
there exist 
two non-negative integers $\delta_S$ and $\delta_W$ such that 
\begin{equation} 
  \label{rem_1} 
  \left(1-T^{\delta_S}_{p_1,\dots,p_n}\right) \Delta_{S}(p_1,\dots,p_n) 
  \,\,=\,\, 0
  \,,~~~~~ 
  \left(1-T^{\delta_W}_{p_1,\dots,p_n}\right) 
  \Delta_{W}(p_1,\dots,p_n,\lambda) 
  \,\,=\,\,0
  \,, 
\end{equation} 
where $T^{\delta}$ is the Taylor subtraction operator;
$\Delta_{S,i}(p_1,\dots,p_n)$ and 
$\Delta_{W,j}( p_1,\dots,p_n,\lambda)$ are the Fourier transformed 
polynomials of $\Delta_{S,i}(x)$ and of $\Delta_{W,j}(x,\lambda)$. 
In power counting renormalizable theories 
the degrees $\delta_S$ and $\delta_W$
are independent of the loop order $n$. 
 
Acting with the zero-momentum subtraction operator 
($1-T^\delta_{S/W}$) on both 
sides of Eqs.~(\ref{qap_1a}) and~(\ref{qap_1b}) leads to 
\begin{eqnarray} 
  \label{delta} 
  (1-T^{\delta_S}) \Big[{\cal S}(\Gamma)\Big]^{(n)} 
  &=& 
  ( 1-T^{\delta_S} ) \Delta^{(n)}_S \,\,=\,\,0
  \,, 
  \nonumber \\ 
  (1-T^{\delta_W}) {\cal W}_{(\lambda)}\left(\Gamma\right)^{(n)} 
  &=&  
  ( 1-T^{\delta_W} ) \Delta^{(n)}_W(\lambda) \,\,=\,\,0
  \,. 
\end{eqnarray} 
Thus the $\Delta^{(n)}$'s are subtracted away. 
The {\it functional} 
Taylor subtraction operator $T^{\delta}$ is defined in (\ref{definitionT2}) 
of Appendix~\ref{app:linear}. 
At the moment we assume that the zero-momentum subtraction is 
possible. This means that the vertex functions have to be sufficiently 
regular at zero momenta. 
In the presence of massless particles, 
zero-momentum subtractions 
of the regularized function $\Gamma^{(n)}$ might lead to 
IR divergences. A practical solution of this problem 
is discussed in~\cite{amt_2}.  
In particular, the decomposition of ${\cal S}_0$  
is made in order to disentangle the IR divergent Green functions from 
the IR finite ones. 
Consequently one has to analyze the new breaking terms arising from 
the commutator between this new operator and the STI and
the WTI operator. Moreover, one can take advantage of the fact that 
the breaking terms are IR safe for general reasons 
provided there are no IR anomalies in the model. 

The l.h.s. of Eqs.~(\ref{delta}) has not yet the correct 
form. Actually, our aim is 
to obtain new STIs and WTIs for the subtracted Green functions, 
i.e. for 
$\gh^{(n)} = (1-T^{\delta_{pc}}) \g^{(n)}$, where $\delta_{pc}$ is the 
naive power counting degree. 
Generically, we have $\delta_W$,$\delta_S  \geq \delta_{pc}$. 
For that purpose 
we commute the Taylor operation with the Slavnov-Taylor 
operator ${\cal S}_0$ and with the 
Ward-Takahashi operator ${\cal W}_{(\lambda)}$
where it is convenient to adopt the decomposition~(\ref{qap_new_1}) 
into a linearized operator plus bilinear terms for the STIs case. 
The part involving the linearized operator leads to 
\begin{eqnarray} 
  (1-T^{\delta_S}) {\cal S}_0(\Gamma^{(n)}) &=& {\cal S}_0 
  \left( \gh^{(n)}\right)+ 
  {\cal S}_0 \left( T^{\delta_{pc}} \Gamma^{(n)} \right) - T^{\delta_S} 
  \left( {\cal S}_0(\Gamma^{(n)}) \right) 
  \,, 
  \label{oversubtraction_1} 
\end{eqnarray} 
and, correspondingly, for the WTIs case, we have 
\begin{eqnarray} 
  (1-T^{\delta_W}) {\cal W}_{(\lambda)}(\Gamma^{(n)})&=&{\cal W}_{(\lambda)} 
  \left( \gh^{(n)}\right)+ 
  {\cal W}_{(\lambda)} \left( T^{\delta_{pc}} 
  \Gamma^{(n)} \right) - T^{\delta_W} 
  \left( {\cal W}_{(\lambda)}(\Gamma^{(n)}) \right) 
  \,. 
  \label{oversubtraction_2} 
\end{eqnarray} 
These equations express the fact that ${\cal S}_0$ and 
${\cal  W}_{(\lambda)}$ are in general not homogeneous in the fields. In 
particular, this is the case for theories with spontaneous symmetry 
breaking.  Notice that, although the Taylor operator is 
scale-invariant, 
it does not commute with ${\cal S}_0$ and ${\cal  W}_{(\lambda)}$ of the SM. 
The difference between $\delta_W,\delta_S$ 
and $\delta_{pc}$ leads to over-subtractions in $\g^{(n)}$. Therefore, 
breaking terms generated by the last two terms on the r.h.s. of 
Eqs.~(\ref{oversubtraction_1}) and~(\ref{oversubtraction_2}) 
have to be introduced.  Furthermore, the action of the 
Taylor operator can be split into the naive contribution 
$\sum_{j=1}^{n-1}\left( {{\rm I}\!\Gamma}^{(j)},{{\rm 
      I}\!\Gamma}^{(n-j)} \right)$ plus the local terms obtained by 
Taylor expansion. These local terms also contribute to the new 
breaking terms. 
 
Finally, by applying the Taylor operator on~(\ref{qap_1a}) 
and using~(\ref{delta}) and~(\ref{oversubtraction_1}) we obtain 
\begin{eqnarray}
  \label{rea} 
  (1 - T^{\delta_S}) \Big[{\cal S}(\Gamma)\Big]^{(n)} &=& 
  {\cal S}_0\left[ (1-T^{\delta_{pc}}) \Gamma^{(n)}\right] + 
  \sum_{j=1}^{n-1}\left( 
  {{\rm I}\!\Gamma}^{(j)},{{\rm I}\!\Gamma}^{(n-j)} 
  \right) 
  \nonumber\\&&\mbox{} 
  -\Big[ T^{\delta_S} {\cal S}_0 - {\cal S}_0 T^{\delta_{pc}}\Big] 
  (\Gamma^{(n)}) - 
  T^{\delta_S}\sum_{j=1}^{n-1}\left( 
  {{\rm I}\!\Gamma}^{(j)},{{\rm I}\!\Gamma}^{(n-j)} 
  \right) \,\,=\,\,  0 
  \,. 
  \nonumber\\
\end{eqnarray} 
The terms in the second line of~(\ref{rea}) represent 
the new local breaking terms which correspond to the 
subtracted function at the $n^{\rm th}$ order, 
$\gh\ton=(1-T^{\delta_{pc}}) \Gamma^{(n)}$. 
Thus, it is convenient to define the new breaking terms as 
\begin{eqnarray} 
  \Psi^{(n)}_S &=&
  \Big[ T^{\delta_S} {\cal S}_0 - {\cal S}_0 T^{\delta_{\rm 
      pc}}\Big](\Gamma^{(n)}) + 
  T^{\delta_S}\sum_{j=1}^{n-1}\left( 
    {{\rm I}\!\Gamma}^{(j)},{{\rm I}\!\Gamma}^{(n-j)} 
  \right) 
  \,. 
  \label{newterms_1} 
\end{eqnarray} 
We emphasize that they are universal in the sense that they do not depend 
on the specific regularization used in the calculation --- in contrast to 
$\Delta^{(n)}_S$ in Eq.~(\ref{qap_1a}) because they are UV-finite. 

Applying the same steps to the WTIs, we end up 
with
\begin{eqnarray} 
  \Psi^{(n)}_W(\lambda) &=& \Big[ T^{\delta_W} {\cal W}_ {(\lambda)}- 
  {\cal W}_ {(\lambda)}  
  T^{\delta_{\rm pc}}\Big] (\Gamma^{(n)}) 
  \,. 
  \label{newterms_2} 
\end{eqnarray} 
As a consequence of the linearity of the WTIs,  Eq.~(\ref{newterms_2}) 
is simpler than Eq.~(\ref{newterms_1}). The form of the 
new breaking terms $\Psi^{(n)}_W(\lambda) $ does not depend on the 
loop-order. In addition, the only
source for $\Psi^{(n)}_W(\lambda)$ 
essentially consists in the different UV behaviour of each term in the 
functional operator ${\cal W}_{(\lambda)}$, 
which lead to over-subtractions. 
The sole source for 
$\Psi^{(n)}_W(\lambda)$ is the spontaneous symmetry-breaking 
mechanism. 

In the described procedure
the use of the Taylor expansion is motivated by practical considerations. 
In momentum space the technique for evaluating Green functions 
with zero external momenta is quite elaborated and, in some cases,
the corresponding integrals can even be solved analytically at the 
three-loop order~\cite{Ste00}. 
 
The algebraic problem is now reduced to finding the proper 
counterterms $\Xi\ton$ which solve the equations 
\begin{eqnarray} 
  {\cal S}_0\left(\Xi^{(n)}\right) = - \Psi^{(n)}_S\,,~~~~~  
  {\cal W}_{(\lambda)}\left(\Xi^{(n)}\right) = - \Psi^{(n)}_W 
  \label{counterterms} 
  \,. 
\end{eqnarray} 
Finally, in terms of $\Xi\ton$  the final correct vertex function at 
$n^{\rm th}$ order reads  
\begin{eqnarray} 
  {{\rm I}\!\Gamma}^{(n)} &=& \gh^{(n)} + \Xi^{(n)} \,\,=\,\, 
  \left(1-T^{\delta_{\rm pc}}\right)\Gamma^{(n)} + \Xi^{(n)}    
  \,. 
\label{final} 
\end{eqnarray} 
 
In order to illustrate the discussion of this subsection, 
we want to present a simple example. We 
consider the renormalization of the three-point function 
$\g^{(n)}_{\hat{W}^{+}_{\nu} \bar{q} b}$.
This amplitude is necessary in the discussion of the
process $b \rightarrow s \gamma$ which is presented in
Section~\ref{sub:application}. 
Assuming that at order $n-1$ 
the renormalization has already been worked out, 
the amplitude  $\g^{(n)}_{\hat{W}^{+}_{\nu} \bar{q} b}$ satisfies the WTI 
\begin{eqnarray} 
  \label{ex_1} 
  && i \left(p_q+p_b\right)_\nu \g^{(n)}_{\hat{W}^{+}_{\nu} 
     \bar{q} b}(p_q,p_b) 
  + i\, M_W  \g^{(n)}_{\hat{G}^{+} \bar{q} b}(p_q,p_b) 
  \nonumber\\&&\mbox{} \hspace{1cm} + 
  \frac{i e }{s_W \sqrt{2}} 
  \left[ \g^{(n)}_{\bar{q} q^\prime}(-p_q)  V_{q^\prime b} P_L 
  -  V_{q q^\prime} P_R \g^{(n)}_{\bar{q}' b}(p_b) 
  \right] \,\,=\,\, \Delta^{(n)}_{W, \lambda^+\bar{q} b}(p_q,p_b) 
  \,, 
\end{eqnarray} 
where $q=u,c,t$ and the sum over $q^\prime$ is understood.
After adopting a renormalization scheme the $n$-loop Green functions
and thus the breaking term 
$\Delta^{(n)}_{W, \lambda^+\bar{q} b}$ are computed.
According to Eq.~(\ref{newterms_2}) we get after 
application of the zero momentum subtraction  
$(1-T^1_{p_q,p_b})$
\begin{eqnarray} 
  i \left(p_q+p_b\right)_\nu 
  \left[ \left(1-T^0_{p_q,p_b}\right) \g^{(n)}_{\hat{W}^{+}_{\nu} 
  \bar{q} b}(p_q,p_b) \right]  
  \hphantom{xxxxxxxxxxxxxxxxxxxxxxxx}
  \nonumber\\ 
  + i\, M_W  \left[ \left(1-T^0_{p_q,p_b}\right) \g^{(n)}_{\hat{G}^{+} 
  \bar{q} b}(p_q,p_b) \right]  
  \hphantom{xxxxxxxxxxxxxxxxxxxxxxxx}
  \nonumber\\ 
  + 
  \frac{i \, e }{s_W \sqrt{2}} 
  \left\{ \left[ (1-T^1_{p_q}) \g^{(n)}_{\bar{q} q'}(-p_q) \right]  
  V_{q^\prime b} P_L 
  -  V_{q q^\prime} P_R \left[(1-T^1_{p_b}) \g^{(n)}_{\bar{q}^\prime b}(p_b) 
  \right] \right\} 
  &=&  
  \nonumber\\ 
  i M_W \Big( p_q \partial_{p_q} + p_b \partial_{p_b} \Big) 
  \left. \g^{(n)}_{\hat{G}^{+} \bar{q} b}(p_q,p_b) 
  \right|_{p_b=p_q=0}  
  &\equiv& 
  \Psi^{W,(n)}_{\lambda^+\bar{q} b}(p_q,p_b)  
  \,,
  \nonumber\\ 
  \label{ex_2} 
\end{eqnarray} 
where we kept all quark masses different from zero.
In the square brackets, the Taylor-subtracted Green function are 
collected. On the r.h.s.
only one finite Green function\footnote{The case where some masses are 
  set to zero and IR divergences occur are discussed in detail in 
  \cite{amt_2}.}  
evaluated for zero external momenta appears. It has been generated by the 
over-subtraction and constitutes
the new breaking term $\Psi^{W,(n)}_{\lambda^+\bar{q} b}$.
Notice that, in contrast to 
$\Delta^{(n)}_{W, \lambda^+\bar{q} b}$, the expression of 
$\Psi^{W,(n)}_{\lambda^+\bar{q} b}$ is  
UV finite quantity and only depends on one Green
function.

In our example the breaking term
$\Psi^{W,(n)}_{\lambda^+\bar{q} b}$ is absorbed by 
the counterterm 
(cf. Eq.~(\ref{cc_8}) of Appendix~\ref{app:cou})
\begin{eqnarray}
  \label{ex_3} 
  \Xi^{W,(n)}_{2} &=&
  \int {\rm d}^4x 
  \left(  \xic{\bar q b W,L}^{W,(n)} \, \bar q \not\!{\hat W}^{+} P_L b   + 
  \xic{\bar q b W,R}^{W,(n)} \, \bar q \not\!{\hat W}^{+} P_R \,b  \right) 
  \,,
\end{eqnarray} 
where the parameters 
$\xic{\bar q b W,L}^{W,(n)}$ and $ \xic{\bar q b W,R}^{W,(n)}$ 
have to be tuned and the wave function renormalization for the quarks
have to be adjusted.
In Section~\ref{sub:application} the above equations will be needed at
one-loop order, i.e. for $n=1$.

Besides the obvious advantages of the Taylor expansion around zero momenta 
--- which essentially  
reduces the number of non-vanishing contributions to the $\Delta\ton$'s 
(for instance, mass counterterms are not needed to restore the STIs or 
WTIs) ---  
there remain complicate expressions for the ``universal'' breaking terms 
$\Psi^{(n)}_S$ as given in Eqs.~(\ref{newterms_1}). Mainly, there are terms 
that depend on lower-order Green functions, recalling the non-linear 
nature of the BRST symmetry. 
Therefore, the WTIs of the background gauge invariance
have to be exploited completely in order to
further increase the practical gain of our method 
with respect to the conventional approach.
In fact, as we will show below, because of algebraic relations 
between the functional operators  
${\cal S}_0$ and ${\cal W}_{(\lambda)}$, restoring the WTIs implies a 
partial restoration also of the STIs.  
This will significantly simplify the evaluation of $\Psi^{(n)}_S$. 
A further simplification in the calculation of the universal 
breaking terms, $\Psi^{(n)}_S$ and $\Psi^{(n)}_W$,  
can be achieved by a suitable choice of the renormalization conditions. 
 
 
\subsection{Computation of counterterms} 
\label{sub:strategy} 
 
The main new aspect in the 
present analysis is 
the BRST variation of the background fields. 
In combination with the 
advantages of the intermediate renormalization, we are able to 
simplify the procedure of the algebraic renormalization. However, an 
essential ingredient for an efficient evaluation of counterterms $\Xi\ton$ is 
the hierarchical structure among functional identities. This 
implies that we can restore the identities one after 
the other without spoiling 
those that  are already recovered. 
Organizing  the counterterms in such a way that 
in a first step the WTIs and in a second one the STIs can be restored 
can only be achieved if the fixing of the STIs does not destroy 
the already restored WTIs. This means that the counterterms needed to 
restore the STIs must be invariant under the action of the 
WTIs, i.e.~they must be background 
gauge invariant. Clearly, this is only possible 
if the breaking terms of the STIs are background gauge invariant. 
The latter is a consequence of consistency conditions between  
the breaking terms of STIs and of WTIs. 
 
In fact, we have to recall that the 
operators ${\cal S }_{\gg }$ and ${\cal W} _{(\lambda)}$ form an 
algebra\footnote{Here we used the notation 
  $(\lambda\wedge\beta)^a=f^{abc}\lambda^b\beta^c, 
  (\lambda\wedge\beta)^A=f^{A+-}\lambda^+\beta^-, \dots$. 
  Notice in addition, that 
  in the present section we are only dealing 
  with the WTIs and STIs. However, our considerations are easily 
  extended to the complete set 
  of functional identities of the SM, as will 
  be recalled in Appendix~\ref{app:ghost_equations}.
}~\cite{brs,STII,grassi}
\begin{eqnarray}\label{cc.1} 
  {\cal S }^2_{\gg }= 0 \hspace{.4cm} {\rm if} \hspace{.4cm}  {\cal 
    S }(\gg )&=&0 \,, 
  \nonumber \\ 
  {\cal S }_{\gg }\left({\cal W} _{(\lambda)}(\gg )\right) 
  - {\cal W} _{(\lambda)}\Big({\cal S }(\gg )\Big) &=&0 
  \,, 
  \nonumber\\ 
   {\cal W} _{(\lambda)} \Big({\cal W} _{(\beta)}(\gg )\Big) - 
  {\cal W} _{(\beta)} \Big({\cal W} _{(\lambda)}(\gg )\Big) &=& 
  {\cal W} _{(\lambda \wedge \beta)}(\gg ) 
  \,, 
\end{eqnarray} 
which, applied to the breaking terms 
$\Psi\ton_S$ and $\Psi\ton_W(\lambda)$ in Eqs.~(\ref{newterms_1}) 
and~(\ref{newterms_2}), leads to the 
so-called Wess-Zumino consistency conditions 
\begin{eqnarray} 
  {\cal W} _{(\lambda)} \Big(\Psi\ton_W(\beta)\Big) - 
  {\cal W} _{(\beta)} \Big(\Psi\ton _W(\lambda)\Big) &=& 
  \Psi\ton_W(\lambda \wedge \beta) 
  \,, 
  \label{cc.3.1} \\ 
  {\cal S} _0 \Big(\Psi\ton_W(\lambda)\Big) 
  - 
  {\cal W} _{(\lambda)}\Big(\Psi\ton_S\Big)  
  &=&0  
  \label{cc.3.2} 
  \,, \\ 
  {\cal S} _0 (\Psi\ton_S) &=& 0 
  \,. 
  \label{cc.3.3} 
\end{eqnarray} 
From well-established results of the cohomology 
analysis of the Wess-Zumino consistency conditions (\ref{cc.3.1}) for 
the functional operators ${\cal W}_{(\lambda)}$~\cite{brs,STII,coho}, 
we know that, in the absence of anomalies, 
the counterterms $\Xi^{W,(n)}$ can be constructed such that 
\begin{eqnarray} 
  {\cal W} _{(\lambda)} \Big(\Xi^{W,(n)}\Big) &=& -\Psi\ton_W(\lambda)  
  \,. 
  \nonumber 
\end{eqnarray} 
Clearly, since ${\cal W} _{(\lambda)}$ 
has a non-trivial kernel, 
this equation cannot fix background gauge invariant counterterms.  
A suitable choice of the latter 
and the organization of the complete set of breaking terms according to the 
quantum numbers of the fields and anti-fields significantly
simplify the computation of $\Xi^{W,(n)}$, as is shown below. 

By adding the counterterms $\Xi^{W,(n)}$ to the Green functions 
$\gh\ton$, we see that 
the WTIs are restored and the STIs are consequently modified 
\begin{eqnarray} 
  \label{eq:app_D_6} 
 {\cal W} _{(\lambda)} \Big(\gh\ton  
  + \Xi^{W,(n)}\Big) &=& \Psi\ton_W(\lambda)  + 
 {\cal  W} _{(\lambda)} \Big(\Xi^{W,(n)}\Big) \,\,=\,\, 0 
  \,, 
 \nonumber\\ 
  \Big[{\cal S} (\gh + \Xi^{W})\Big]\ton &=& \Psi\ton_S +  {\cal S} _0\Big( 
  \Xi^{W,(n)}\Big) \,\,\equiv\,\, \hat{\Psi}\ton_S 
  \,. 
 \nonumber 
\end{eqnarray} 
Moreover, as a consequence of Eq.~(\ref{cc.3.2}), 
the new breaking terms $\hat{\Psi}\ton_S $ are explicitly background 
gauge invariant as can be seen as follows:
\begin{eqnarray}\label{eq:D_7} 
  {\cal W} _{(\lambda)} \Big( \hat{\Psi}\ton_S \Big) ={\cal W}
  _{(\lambda)}\Big( \Psi\ton_S\Big)  
  +   {\cal W} _{(\lambda)} {\cal S}_0 \Big( 
  \Xi^{W,(n)}\Big) = {\cal W} _{(\lambda)}\Big(\Psi\ton_S\Big) -  {\cal S} _0 
  \left(  \Psi\ton_W(\lambda)\right) = 0\,. 
  \nonumber 
\end{eqnarray} 
The difference between a conventional approach without the BFM 
and our method is essentially that 
the number of independent breaking terms $\hat{\Psi}\ton_S$ is significantly
reduced. Indeed, only background 
gauge invariant counterterms\footnote{In Appendix~\ref{app:cou} 
  the complete structure of the counterterms is given.}
$\Xi^{S,(n)}$ are needed 
to restore non-linear STIs~(\ref{newterms_2})
\begin{eqnarray} 
  \label{eq:D_8} 
 {\cal W} _{(\lambda)} \Big(\Xi^{S,(n)}\Big) = 0\,, ~~~~~ 
 {\cal S} _0\Big(\Xi^{S,(n)}\Big)= - \hat{\Psi}\ton_S\,. 
\end{eqnarray} 

However, this is not the total benefit we obtain from the inclusion of the
BFM. It is also possible to simplify the calculation of 
the second equation in~(\ref{eq:D_8}).
To this end we decompose the counterterms 
$\Xi^{S,(n)}$ according to the dependence on the anti-fields
$\phi^*$ and on the fields $\Omega$
\begin{eqnarray} 
  \Xi^{S,(n)} &=& \Xi^{S,(n)}_O[\phi, \hat\phi] 
                  + \Xi^{S,(n)}_\#[\phi,\hat\phi,\phi^*,\Omega]
  \,, 
  \nonumber \\ 
  \Xi^{S,(n)}_\#[\phi,\hat\phi,\phi^*,\Omega] 
               &=& \Xi^{S,(n)}_{\#,O}[\phi,\hat\phi,\phi^*] 
                   + \Xi^{S,(n)}_\Omega[\phi^*,\Omega]
  \,, 
  \label{eq:D_9} 
\end{eqnarray} 
where $\Xi_O^{S,(n)}[\phi, \hat\phi]$ depends only on quantum and background
fields. Note that due to power counting the dependence of
$\Xi^{S,(n)}_\#[\phi,\hat\phi,\phi^*,\Omega]$
(and thus also of $\Xi_\Omega^{S,(n)}[\phi^*,\Omega]$) 
on $\phi^*$ and $\Omega$ is only linear.
For the same reason 
$\Xi_\Omega^{S,(n)}[\phi^*,\Omega]$ does not depend on $\phi$ and
$\hat{\phi}$. 
The separation of $\Omega$-dependent and $\Omega$-independent terms in the
second line of Eq.~(\ref{eq:D_9}) turns out to be useful as we will
see below.
In addition, the background gauge invariance of $\Xi^{S,(n)}$ implies
that each contribution to the  
r.h.s. of Eq.~(\ref{eq:D_9}) is a linear combination of invariant polynomials.

On the other hand, the background gauge invariant breaking terms
$\hat{\Psi}\ton_S$  
can also be decomposed according to the occurrence of $\phi^*$ and
$\Omega$,
\begin{eqnarray} 
  \label{eq:D_10} 
  \hat{\Psi}\ton_S &=& \hat{\Psi}\ton_{S,O}[\phi, \hat\phi] 
               +\hat{\Psi}\ton_{S,\#}[\phi,\hat\phi,\phi^*] 
               + \hat{\Psi}\ton_{S,\Omega O}[\phi,\hat\phi,\Omega]
               + \hat{\Psi}\ton_{S,\Omega\#}[\phi,\hat\phi,\phi^*,\Omega]
  \,,
\end{eqnarray} 
where the corresponding dependence is again linear.

Our final aim is to re-write the second equation of~(\ref{eq:D_8})
using the decompositions introduced in~(\ref{eq:D_9}) and~(\ref{eq:D_10}).
In this respect the following definitions
turn out to be useful
\begin{eqnarray} 
  \label{de_4} 
  X^0_{\phi}(x) \,\,\equiv\,\, 
  \left. \frac{\delta \g_0}{\delta \phi(x)} \right|_{\phi^*=0}
  \,,~~~~~~&&
  X^\#_{\phi}(x) \,\,\equiv\,\, 
  \frac{\delta \g_0}{\delta \phi(x)} - X^0_{\phi}(x)
  \,, 
  \nonumber 
\end{eqnarray} 
where $\g_0$ is the classical reduced functional.
$X^0_{\phi}$ coincides with 
the classical equations of motion of the field $\phi$
and $X^\#_{\phi}$ is the anti-field part of the equations of motion. 
It is also useful to introduce the compact notation
\begin{eqnarray} 
  \label{de_5} 
  \sum_\phi \int {\rm d}^4x \, 
  \Omega\left( \frac{\delta}{\delta \hat \phi} 
    - \frac{\delta}{\delta \phi}\right)  
  &\equiv& 
  \int {\rm d}^4x \, \sum_{A =3,\pm,a} 
  \Omega^{\mu A} \left( \frac{\delta}{\delta \hat V^{A\mu}} 
    - \frac{\delta}{\delta V^{A\mu}}\right) 
  \nonumber\\&&\mbox{}
  + 
  \int {\rm d}^4x \, \sum_{i=H,0,\pm} 
  \Omega^{i} \left( \frac{\delta}{\delta \hat \Phi_i} 
    - \frac{\delta}{\delta  \Phi_i}\right)
  \,. 
  \nonumber 
\end{eqnarray} 
We can then write Eq.~(\ref{eq:D_8}) using Eqs.~(\ref{ST}),
(\ref{line_ST}) and~(\ref{bracket})
in the following schematic way
\begin{eqnarray}
  \sum_\phi \int {\rm d}^4x
  \left[
  \frac{\delta\Gamma_0}{\delta\phi}
  \frac{\delta\Xi^{S,(n)}}{\delta\phi^*}
  +
  \frac{\delta\Xi^{S,(n)}}{\delta\phi}
  \frac{\delta\Gamma_0}{\delta\phi^*}
  + \Omega\left(\frac{\delta}{\delta\hat{\phi}}
               -\frac{\delta}{\delta\phi}
          \right) \Xi^{S,(n)}
  \right]
  &=& \hat{\Psi}_S^{(n)}
  \,.
  \nonumber
\end{eqnarray}
In this equation we insert Eqs.~(\ref{eq:D_9}) and~(\ref{eq:D_10})
and sort the terms according to their dependence
on $\phi^*$ and $\Omega$. This leads to the system of equations
\begin{eqnarray} 
  \label{de_6} 
  \sum_\phi \int {\rm d}^4 x  \left(  s\, \phi \,\frac{\delta }{\delta \phi}
    + X^\#_{\phi}\frac{\delta }{\delta \phi^*} \right) \Xi^{S,(n)}_{\#,O} 
  &=& 
  \hat{\Psi}\ton_{S,\#} 
  \,,
  \nonumber\\ 
  \sum_{\phi}  \int {\rm d}^4 x \left[X^\#_{\phi} \frac{\delta}{\delta
  \phi^*}  \Xi^{S,(n)}_\Omega + 
  \Omega\left( \frac{\delta}{\delta \hat \phi} - 
    \frac{\delta}{\delta \phi}\right)  \Xi^{S,(n)}_{\#,O} \right]
  &=&  
  \hat{\Psi}\ton_{S,\Omega \#}
  \,,
  \nonumber\\  
  \sum_{\phi} \int {\rm d}^4 x \left[ X^0_{\phi} \frac{\delta}{\delta \phi^*}  
  \Xi^{S,(n)}_\Omega + \Omega\left( \frac{\delta}{\delta \hat \phi} 
  - \frac{\delta}{\delta \phi}\right)  \Xi^{S,(n)}_O \right] 
  &=&  
  \hat{\Psi}\ton_{S,\Omega O}
  \,,
\end{eqnarray} 
which can be solved for the counterterms
$\Xi^{S,(n)}_O$, $\Xi^{S,(n)}_{\#,O}$ and $\Xi^{S,(n)}_\Omega$. 
For completeness we also list the equation for
$\hat{\Psi}\ton_{S,O}$ which reads
\begin{eqnarray}
  \label{de_6_1} 
  \sum_\phi \int {\rm d}^4 x \left[
    s\, \phi \,\frac{\delta }{\delta \phi} \Xi^{S,(n)}_{O}
    + X^0_{\phi}\frac{\delta }{\delta \phi^*} \Xi^{S,(n)}_{\#,O} 
  \right]
  &=& 
  \hat{\Psi}\ton_{S,O} 
  \,.
\end{eqnarray}
In principle Eq.~(\ref{de_6_1}) could be combined with the first
equation of~(\ref{de_6}) in order to obtain $\Xi^{S,(n)}_{O}$. 
However, the resulting expressions become more complicated and it is
more advantageous to use only the system~(\ref{de_6}).

In the derivation of Eqs.~(\ref{de_6}) and~(\ref{de_6_1}) we have used
\begin{eqnarray}
  \frac{\delta\Gamma_0}{\delta\Omega} \,\,=\,\,0
  \,,
  &&
  \frac{\delta\Gamma_0}{\delta\phi^*} \,\,=\,\,s\,\phi
  \,,
  \label{eq:auxeq}
\end{eqnarray}
where $s\,\phi$ is the classical BRST transformation of $\phi$.
The first equation is a consequence of the fact that we used the
reduced functional and
the second equation in~(\ref{eq:auxeq}) 
immediately follows from~(\ref{gree.5}).

At this point some comments on~(\ref{de_6}) are in order.
The big advantage of our approach is that 
conventional STIs obtained by differentiating Eq.~(\ref{ST}) with 
respect to one ghost field $c^B$ and some quantum fields
are completely avoided.
More precisely, it is not necessary to evaluate the breaking terms 
$\hat\Psi\ton_{S,O}[\phi,\hat{\phi}]$ which enormously reduces the
calculational effort.
The complete set of counterterms 
$\Xi^{S,(n)}_O$ are fixed in terms of background Green functions
by the extension of the BRST symmetry 
obtained by including the background fields.
A further simplification 
is due to the power counting degrees of $\Omega$ and their Lorentz 
properties. The needed STIs are simpler than those obtained by 
differentiating with respect to a ghost field $c^B$. 

Notice that, although we can completely avoid the STIs for quantum
fields, we cannot totally skip the STIs obtained with at least one
anti-field, namely the first equation of system~(\ref{de_6}). However,
this requires some remarks. The reason for this has to be ascribed to
the difference between BRST symmetry and  gauge symmetry. Indeed, the
anti-field-dependent terms are BRST-invariant and not gauge invariant.
This means that in the gauge invariant part, the fields $\phi$ and
$\hat\phi$ appear only in the combination 
$Z_\phi \phi + \hat\phi$, while in the
anti-field-dependent terms this is not valid 
and, in principle, all the possible combinations
could appear (cf. Appendix~\ref{app:cou}).
Nevertheless, the background gauge invariance already removes some of
the breaking terms $\hat{\Psi}\ton_{S,\#}$, leaving only few free
parameters. This will be explicitly shown in the forthcoming section.

At this point we should spend some words on the computation of the
breaking terms 
$\hat{\Psi}^{(n)}_{S,\#}$,
$\hat{\Psi}^{(n)}_{S,\Omega O}$
and
$\hat{\Psi}^{(n)}_{S,\Omega\#}$
which appear on the r.h.s. of Eq.~(\ref{de_6}).
Guided by 
the structure of the second and third equations of~(\ref{de_6})
it appears useful to consider relations between the Green functions
$\g_{\phi_1\phi_2\ldots}$
and
$\g_{\hat\phi_1\phi_2\ldots}$ 
where one quantum field is replaced by the corresponding background
field.
Such relations, that contain $\hat{\Psi}^{(n)}_{S,\Omega O}$
and $\hat{\Psi}^{(n)}_{S,\Omega\#}$ as breaking term,
can be derived as follows:
\begin{itemize}
  \item 
  Substitute $\hat\phi_1$ by $\Omega_1$ (the BRST variation of
  $\hat\phi_1$) and take the derivative of the STI~(\ref{ST}) 
  with respect to $\Omega_1$ and $\phi_2\ldots$ 
  \begin{eqnarray}
  \label{rule_1}
  \left. \frac{\delta {\cal S}(\g) }
              {\delta \Omega_1(p_1) 
               \delta \phi_2(p_2) \ldots }\right|_{\phi=0}
  &=&  
  \g_{\hat\phi_1 \phi_2 \ldots } 
  - 
  \g_{\phi_1 \phi_2 \ldots } 
  + \dots
  \nonumber\\
  &=& \Delta_{S,\Omega_1\phi_2\ldots}
  \,, 
  \end{eqnarray}
  where the ellipses contain those terms which are quadratic in $\g$. 

  \item 
  Recall that $\Omega$ carries a Faddeev-Popov charge $+1$. Thus
  on the r.h.s. of Eq.~(\ref{rule_1})
  only the Green functions with vanishing ghost charge are different 
  from zero. This, together with the Lorentz invariance, 
  selects non-vanishing contributions from the ellipses in~(\ref{rule_1}). 
\end{itemize}
Concerning the extraction of identities for $\hat{\Psi}^{(n)}_{S,\#}$
(see the first equation of~(\ref{de_6})) we refer to the
detailed discussion presented in Sections~2.3 and~3.2 of Ref.~\cite{amt_1}.

The existence of a solution to the system~(\ref{de_6}) is guaranteed by
the Wess-Zumino consistency conditions~(\ref{cc.3.1}) and~(\ref{cc.3.2}).
The solution is not unique since the Slavnov-Taylor operator ${\cal S}_0$
has a non-trivial kernel. This means that there exist background gauge
invariant polynomials
that satisfy ${\cal S}_0 \left( \Xi^{N,(n)} \right) = 0$. In addition, also
the 
invariant counter\-terms $\Xi^{N,(n)}$ can be split 
into an anti-field-independent part
$\Xi^{N,(n)}_O[\phi,\hat\phi]$ and
an anti-field-dependent one $\Xi^{N,(n)}_\#[\phi,\hat\phi,\phi^*,\Omega]$.
From this decomposition, it is straightforward to see that the
anti-field-dependent
counter\-terms  $\Xi^{N,(n)}_\#[\phi,\hat\phi,\phi^*,\Omega]$ parameterize the
unphysical normalization conditions (such as 
wave function and gauge parameter renormalizations). 
The invariant counter\-terms
$\Xi^{N,(n)}_O[\phi,\hat\phi]$ turn
out to depend upon  linear combinations
$\tilde\phi = Z_\phi \phi + \hat\phi$ (this can be seen by solving
the homogeneous system (\ref{de_6})). They parameterize the physical
normalization
conditions that can be employed to match the physical
parameters.

Having solved  Eqs.~(\ref{de_6}), the result of the algebraic 
problem~(\ref{qap_new_2}) can be summarized into the following 
equations:  
\begin{eqnarray} 
  \label{de_7} 
  \gg\ton &=& \g\ton + \g^{CT,(n)}
  \,,  
  \nonumber \\ 
  \g^{CT,(n)} &=& -T^{\delta_{pc}}\g\ton + \Xi^{W,(n)}  
                  + \Xi^{S,(n)} + \Xi^{N,(n)} 
  \,. 
\end{eqnarray} 
The last term, $\Xi^{N,(n)}$, is an invariant counterterm and has to
be added in order to tune the normalization conditions on physical data. 
 
Before discussing the more complex example of $b\to s\gamma$ in 
Section~\ref{sub:application}, we would like to consider as a 
simple case the gluon two-point function. 
In this example it is assumed that the breaking terms are already
known and background gauge
invariant and Eqs.~(\ref{de_6}) are solved using the decomposition
introduced in Eqs.~(\ref{eq:D_9}) and~(\ref{eq:D_10}). 
 
The final aim is the computation of the counterterms for  
the quantum, the background 
and the corresponding mixed two-point functions denoted by 
$\Xi^{S,(n)}_{V^a_\mu V^b_\nu}(p)$, 
$\Xi^{S,(n)}_{\hat V^a_\mu \hat V^b_\nu}(p)$ 
and 
$\Xi^{S,(n)}_{\hat V^a_\mu V^b_\nu}(p)$, 
respectively. 
In the notation of Eq.~(\ref{eq:D_9}) they correspond to 
$\Xi^{S,(n)}_O$ which occurs in the last equation in~(\ref{de_6}). 
Differentiation with respect to $\Omega^b_\nu$ and $\hat V^a_\mu$ or 
$\Omega^b_\nu$ and $V^a_\mu$ leads to the two equations 
\begin{eqnarray} 
  -\g_{0,\hat V^a_\mu V^c_\rho}(p) \, 
  \Xi^{S,(n)}_{V^{*,c}_\rho \Omega^b_\nu}(p) 
  + \Xi^{S,(n)}_{\hat V^a_\mu \hat V^b_\nu}(p) 
  - \Xi^{S,(n)}_{\hat V^a_\mu V^b_\nu}(p) 
  &=& 
  \hat\Psi^{(n)}_{S, \Omega^b_\nu \hat V^a_\mu}(-p) 
  \,, 
  \nonumber\\ 
  -\g_{0,V^a_\mu V^c_\rho}(p)  \, 
  \Xi^{S,(n)}_{V^{*,c}_\rho \Omega^b_\nu}(p) 
  + \Xi^{S,(n)}_{V^a_\mu \hat V^b_\nu}(p) 
  - \Xi^{S,(n)}_{V^a_\mu V^b_\nu}(p) 
  &=& 
  \hat\Psi^{(n)}_{S, \Omega^b_\nu V^a_\mu}(-p) 
  \,, 
  \label{ex_co.1} 
\end{eqnarray} 
which contain four unknown counterterms. 
These equations can be used to determine 
$\Xi^{S,(n)}_{V^a_\mu V^b_\nu}(p)$ and 
$\Xi^{S,(n)}_{\hat V^a_\mu V^b_\nu}(p)$, as 
$\Xi^{S,(n)}_{\hat V^a_\mu \hat V^b_\nu}(p)$ is fixed by normalization 
conditions for physical parameters~\cite{msbkg}.
In addition they are transverse as a consequence of the WTI.
Furthermore, the counterterm $\Xi^{S,(n)}_{V^{*,c}_\rho \Omega^b_\nu}(p)$ 
is constrained by the second equation of~(\ref{de_6}). 
In particular, differentiating with respect to $c^c$, $\Omega^a_\mu$ and 
$V^{*,b}_\nu$ gives 
\begin{eqnarray} 
  \label{ex_co.3} 
  -\g_{0,V^d_\rho V^{*,b}_\nu c^c}(p,q)\, 
  \Xi^{S,(n)}_{V^{*,d}_\rho \Omega^a_\mu}(-p-q) 
  - \Xi^{S,(n)}_{V^a_\mu V^{*,b}_\nu c^c}(p,q) 
  + \Xi^{S, (n)}_{\hat V^a_\mu V^{*,b}_\nu c^c}(p,q) 
  &=& 
  \hat\Psi^{(n)}_{S, \, \Omega^a_\mu V^{*,b}_\nu c^c}(p,q) 
  \,. 
  \nonumber\\ 
\end{eqnarray} 
This equation contains the new unknown terms 
$\Xi^{S, (n)}_{\hat V^a_\mu V^{*,b}_\nu c^c}$ 
and 
$\Xi^{S,(n)}_{V^a_\mu V^{*,b}_\nu c^c}$. 
The latter also occurs in the first equation of~(\ref{de_6}) after 
taking the derivatives with respect to 
$c^a$, $c^b$ and $V^{*,c}_\nu$ 
\begin{eqnarray} 
  \label{ex_co.4} 
  -\g_{0,c^a V^{*,d}_\rho}(p+q) \Xi^{S,(n)}_{V^d_\rho V^{*,c}_\nu c^b}(q,p) + 
  \g_{0,c^b V^{*,d}_\rho}(-p) \Xi^{S,(n)}_{V^d_\rho V^{*,c}_\nu c^a}(q,-p-q)
  \nonumber\\\mbox{}
  -
  \g_{0,c^a c^b c^{*,d}}(p,q) \Xi^{S,(n)}_{c^d V^{*,c}_\nu}(q) 
  - 
  \g_{0,V^d_\rho V^{*,c}_\nu c^b}(q,p) \Xi^{S,(n)}_{c^a V^{*,d}_\rho}(p+q)
  \nonumber\\\mbox{}
  +
  \g_{0,V^d_\rho V^{*,c}_\nu c^a}(q,-p-q) \Xi^{S,(n)}_{c^b V^{*,d}_\rho}(-p) 
  - 
  \g_{0,c^d V^{*,c}_\nu}(q) \Xi^{S,(n)}_{c^a c^b c^{*,d}}(p,q) 
  &=& 
  \hat\Psi^{(n)}_{S, c^a c^b V^{*,c}_\nu}(p,q) 
  \,. 
  \nonumber\\
\end{eqnarray} 
Thus, after exploiting~(\ref{de_6}) 
we are left with the counterterms 
$\Xi^{S,(n)}_{\hat V^a_\mu V^{*,b}_\nu c^c}$, 
$\Xi^{S,(n)}_{c^a V^{*,d}_\rho}$ 
and 
$\Xi^{S,(n)}_{c^a c^b c^{*,d}}$ 
which are not fixed by the Eqs.~(\ref{ex_co.1}), (\ref{ex_co.3}) 
and~(\ref{ex_co.4}). 
However, due to background gauge invariance the counterterms 
$\Xi^{S,(n)}_{\hat V^a_\mu V^{*,b}_\nu c^c}$ 
and 
$\Xi^{S,(n)}_{c^a V^{*,d}_\rho}$ 
are not independent, rather one has
$i p_\nu \Xi^{S,(n)}_{\hat V^a_\mu V^{*,b}_\nu c^c} =
f^{dbc} \Xi^{S,(n)}_{c^a V^{*,d}_\mu}$. 
Furthermore, 
$\Xi^{S,(n)}_{c^a V^{*,d}_\rho}$ and 
$\Xi^{S,(n)}_{c^a c^b c^{*,d}}$ 
are fixed via normalization conditions, namely the wave function 
renormalization for the quantum 
gauge field $V^a_\mu$ and the one for the ghost field $c^a$, 
respectively\footnote{See also Section~\ref{sub:application} below.}. 
Thus all counterterms appearing in the computation of vector boson 
two-point functions are determined. 
 
Notice that Eqs.~(\ref{ex_co.1}), 
(\ref{ex_co.3}) and (\ref{ex_co.4}) are quite simple and they can 
easily be solved in terms of  $\Xi\ton$. Furthermore, once the breaking 
terms, which are expressed in terms of vacuum integrals, are computed 
also the individual coefficients of the counterterms 
(see Appendix~\ref{app:cou}, Eq.~(\ref{cc_5})) are available. 
In this context we want to stress again that the counterterms 
$\Xi^{S,(n)}_{V^a_\mu V^b_\nu}(p)$, 
$\Xi^{S,(n)}_{\hat V^a_\mu \hat V^b_\nu}(p)$ 
and 
$\Xi^{S,(n)}_{\hat V^a_\mu V^b_\nu}(p)$ 
have to be background gauge invariant.


\setcounter{equation}{0} 
\section{Application} 
\label{sub:application} 

Based on the theoretical analysis of the previous section,
we consider in this section the important process 
$b\rightarrow s \gamma$
(see, e.g., Ref.~\cite{FerYaoOliRay95} where the algebraic method has
been applied to $b\rightarrow s \gamma$ or Refs.~\cite{Greub,GamHai00}
for recent summaries on QCD and electroweak corrections, respectively.)
and formulate a strategy for computing the corresponding amplitudes 
at two-loop order independent of any regularization.
This amplitude is used as an illustrating example in order to show 
how the complete  procedure works in detail. We 
assume that all symmetry identities which are involved are broken 
and have to be restored. However,  we stress again  
that our procedure is independent of the regularization and  
a suitable choice allows for many additional simplifications
within our approach.
Indeed, it is easy to see that many breaking terms discussed in the
following do not occur in a reasonable non-invariant regularization.

Since at the classical level there are no flavour changing neutral currents
(FCNC) in the SM,
the amplitude $\gg_{\hat A_\mu \bar s b}$ vanishes at tree level.
To compute it up to two loops, one has to take  into account only one-loop
counterterms
as the two-loop amplitude --- even if superficially divergent ---
is directly related by the linear WTIs to a superficially convergent
amplitude,
which is completely identified once  the one-loop counterterms are known.
If one had an invariant regularization,  the two-loop amplitude would appear
superficially convergent. However, in the absence of a non-invariant
regularization, the amplitude might turn out to be divergent. Therefore,
there remains the problem of computing the one-loop non-invariant
counterterms in an efficient way.

For the physical decay rate $b\to s\gamma$, in principle, only 
the form factor of the magnetic moment has to be computed.
The discussion in this section, however, is kept more general
and the complete $\hat{A}\bar{b}s$ vertex is considered.

From the topological structure of the two-loop diagrams 
contributing to the amplitude
$\gg^{(2)}_{\hat{A}_{\mu} b \bar{s}}$, 
it is evident that the 1PI three- and four-point functions 
with external gauge or scalar quantum fields 
(e.g. $\gg^{(1)}_{V^{\alpha}_{\mu} V^{\beta}_{\nu} V^{\gamma}_{\rho}}$
or 
$\gg^{(1)}_{V^{\alpha}_{\mu} V^{\beta}_{\nu} V^{\gamma}_{\rho}
V^{\delta}_{\sigma}}$)
or with ghost fields 
(e.g. $ \gg^{(1)}_{V^{\alpha}_{\mu} \bar{c}^{\beta} c^{\gamma}}$) 
do not appear as one-loop sub-graphs.
Actually, the renormalization of sub-divergences 
with more than two 
(gauge or scalar) quantum fields enters the BFM calculations only 
starting from the three-loop order.  
Thus, we only have to consider the three-point functions 
\begin{eqnarray} 
  \label{gr:bsga} 
  \begin{array}{llll} 
    \gg^{(1)}_{\hat{A}_{\mu} \bar{s} b}(p_{s},p_{b})\,, 
    & 
    \gg^{(1)}_{\hat{A}_{\mu} \bar{q}_{2} q_1}(p_{2},p_{1})\,, 
    & 
    \gg^{(1)}_{\hat{A}_{\mu} W^{+}_{\nu} W^{-}_{\rho}} 
    (p_{+},p_{-})\,, 
    & 
    \gg^{(1)}_{\hat{A}_{\mu} G^{+} G^{-}} (p_{+},p_{-})\,, 
    \\ 
    \\ 
    \gg^{(1)}_{\hat{A}_{\mu} G^{\pm} W^{\mp}_{\nu}} (p_{\pm},p_{\mp})\,, 
    & 
    \gg^{(1)}_{\hat{A}_{\mu} \bar c^{\pm} c^{\mp}} (p_{\pm},p_{\mp})\,, 
    & 
    \gg^{(1)}_{W^{\pm}_{\mu} \bar{q}_{2} q_1}(p_{2},p_{1})\,, 
    & 
    \gg^{(1)}_{G^{\pm} \bar{q}_{2} q_1}(p_{2},p_{1})\,, 
  \end{array} 
\end{eqnarray} 
and the two-point functions 
\begin{eqnarray}\label{gr:bsgb} 
  \begin{array}{lll} 
    \gg^{(1)}_{\bar{s} b}(p_{b})\,, 
    & 
    \gg^{(1)}_{\bar{q}_{2} q_1}(p_{1})\,, 
    & 
    \gg^{(1)}_{W^{+}_{\nu} W^{-}_{\rho}} (p_{-})\,, 
    \\ 
    \\ 
    \gg^{(1)}_{G^{+} G^{-}} (p_{-})\,, 
    & 
    \gg^{(1)}_{G^{\pm} W^{\mp}_{\mu}} (p_{\mp})\,, 
    & 
    \gg^{(1)}_{\bar c^{\pm} c^{\mp}} (p_{\mp})\,. 
  \end{array} 
\end{eqnarray} 
In Eqs.~(\ref{gr:bsga}) and ~(\ref{gr:bsgb}) 
$q_1$ and $q_2$ are two generic quark fields. 
Notice that the Green functions 
$\gg_{\hat{A}_{\mu} G^{+} W^{-}_{\nu}}$ and 
$\gg_{\hat{A}_{\mu} W^{+}_{\nu} G^{-}}$ have no 
tree-level contribution (see, e.g.,~\cite{msbkg}). 
At the one-loop level, however, contributions may appear 
as soon as a non-invariant regularization scheme is used. 
 
For the computation we follow the strategy outlined in  
Section~\ref{sub:strategy}. 
In a first step we exploit the WTIs for the background fields and 
fix all possible counterterms. In a second step 
the remaining counterterms are determined by STIs and,  
finally, we tune the free parameters of the theory to match the 
normalization conditions.  
The complete set of counterterms, be they used to restore  
the WTIs ($\Xi^W$) or the STIs ($\Xi^S$), or to implement the  
normalization conditions ($\Xi^N$), can be separated into  
anti-field-dependent counterterms 
$\Xi_\#[\phi,\hat\phi,\phi^*,\Omega]$ 
and anti-field-independent ones 
$\Xi_O[\phi,\hat\phi]$ (cf. Eq.~(\ref{eq:D_9}))  
according to the ghost number.  
In principle the former should be determined first as can be
deduced from 
the triangular structure of Eq.~(\ref{de_6}). 
However, for pedagogical purposes, we discuss the renormalization starting 
directly from the analysis of the amplitudes~(\ref{gr:bsga}) and 
(\ref{gr:bsgb}), i.e. from the last equation in~(\ref{de_6}). 
The anti-field-dependent counterterms turn out to be 
necessary along the discussion. 
Moreover, as mentioned in  
Section~\ref{sub:remove}, the intermediate  
zero-momentum subtraction is applied to simplify the computation.  

In order to provide a guidance for the reader we briefly 
outline the rest of this section. 
We start with the WTIs containing those Green functions  
of our list (see Eqs.~(\ref{gr:bsga}) and~(\ref{gr:bsgb})) 
involving background 
fields and determine the corresponding counterterms. 
In our case this means that we have to consider the identities 
involving Green functions with an external background photon. 
In the corresponding equations two-point Green 
functions with external quantum fields appear. 
According to our procedure this means that in a first step  
the WTIs for the background two-point functions have to be considered 
(cf. point~2 below) and afterwards the STIs for the  
quantum counterparts (cf. point~3) are investigated. 
This completes the determination of the counterterms for the 
background three-point functions in Eq.~(\ref{gr:bsga}). 
In a next step (cf. point~4) the quantum three-point functions  
of~(\ref{gr:bsga}) are discussed. 
To keep the discussion simple we will postpone the treatment of those 
Green functions that do not contribute to the counterterms of the
quantum fields in points~1 to~4. The corresponding identities are
discussed in point~5.
 
\vskip.3cm 
\noindent 
{\it 1. Background three-point functions}\footnote{In 
  the following the reduced functional discussed briefly 
  at the end of Section~\ref{sec:sti} and
  in Appendix~\ref{app:ghost_equations}. 
  This slightly simplifies the forthcoming equations.}.
The main goal of our analysis is to obtain the counterterm for the
amplitude $\g_{\hat{A}_{\mu} \bar{s} b}$
that constitutes the central object in the process $b\to s\gamma$.
In this part we discuss the one-loop amplitudes involving the 
background photon $\hat{A}$ and two fermions or gauge bosons, respectively.

The WTI which contains the amplitude 
$\g^{(1)}_{\hat{A}_{\mu} \bar{q}_{2} q_1}$ reads (cf.~\cite{amt_1}):
\begin{eqnarray} 
  \label{bsg1} 
  \lefteqn{\left. \frac{\delta^3 {\cal W}_{(\lambda)}(\g^{(1)})} 
  {\delta \lambda_A(-p_1 - p_2) \delta \bar{q}_2(p_2)  \delta q_1(p_1)} 
  \right|_{\phi=0} = } 
  \nonumber \\&&\mbox{} 
   i \left(p_{2} + p_{{1}}\right)^\mu \g^{(1)}_{\hat{A}_{\mu} 
  \bar{q}_2 q_1}(p_{2},p_{{1}}) 
  + i e Q_{q} \left(  \g^{(1)}_{\bar{q}_2 q_1}(p_{1})  - 
  \g^{(1)}_{\bar{q}_2 q_1}(-p_2) \right) 
  \,\,=\,\,\Delta_{W,\lambda_A \bar{q}_2 q_1} 
  \,, 
\end{eqnarray} 
where $Q_{q}$ is the common charge of the quarks $q_1$ and $q_2$.
The equation for $\g^{(1)}_{\hat{A}_{\mu} \bar{s} b}$ is obtained
by the replacements $q_2\to s$ and $q_1\to b$.
As the breaking term $\Delta^W_{\lambda_A \bar{q}_2 q_1}$ has 
mass dimension one
we remove them by acting with $(1 - T^1_{p_1,p_2})$.
The resulting Green functions 
\begin{eqnarray}
  \label{eq:sub_1} 
  \gh^{(1)}_{\hat{A}_{\mu} \bar{q}_2 q_1}(p_2,p_1)\,\,=\,\, 
  (1 - T^0_{p_1,p_2}) \g^{(1)}_{\hat{A}_{\mu} \bar{q}_2 q_1}(p_2,p_1) 
  \,,&~~~~&  
  \gh^{(1)}_{\bar{q}_2 q_1}(p)\,\,=\,\, 
  (1 - T^1_p) \g^{(1)}_{\bar{q}_2 q_1}(p) 
  \,, 
\end{eqnarray} 
automatically satisfy the WTI,
which means that the counterterm $\Xi^{W,(1)}[\phi,\hat\phi]$ is zero.
However, we still have the freedom to impose normalization conditions
for the quark self-energies.
In particular, we can add the (BRST and background gauge invariant) 
counterterms 
\begin{eqnarray}
  \label{qua_nor}
  \Xi^{N,(1)}_{1}
  &=&
    \sum_{\psi \psi'}
  \int {\rm d}^4 x
  \Big( \xic{\bar \psi \nabla \psi^\prime}^{N,(1)}
          \bar \psi^I \not\!\nabla_{IJ} \psi^{\prime,J}
      +  \xic{\bar\psi\psi^\prime\Phi,m}^{N,(1)} Y^{i,IJ}_m
          (\Phi +\hat\Phi + v)_i \bar \psi_I \psi'_J  + {\rm h.c.}
  \Big)
  \,,
\end{eqnarray}
where $\xic{\bar \psi \nabla \psi^\prime}^{N,(1)}$ and 
$\xic{\bar\psi\psi^\prime\Phi,m}^{N,(1)}$ are the coefficients of the
counterterms and
$Y^{i,IJ}_m $ are invariant tensors of the 
fermion and scalar representations (cf. Appendix~\ref{app:cou}, 
Eq.~(\ref{cc_8})).
$v^i$ is the vacuum expectation value. 
The free parameters are fixed by normalization 
conditions for the CKM matrix elements\footnote{Notice 
  that the WTI also implies 
  special normalization conditions for the CKM 
  elements as described in~\cite{gg}.} 
and by quark mass renormalizations~\cite{strumia,gambino,gg} 
(see also~\cite{grassi} for a complete discussion of normalization 
conditions within the BFM framework). 
In particular, the diagonal part in Eq.~(\ref{qua_nor}) can be written
as
\begin{eqnarray}
  \Xi^{N,(1)}_{\bar q q}(p)
   &=& 
   \xi^{(1)}_{2,q} \left(\not\!p - m_q \right) 
  + \xi^{(1)}_{q} m_q
  \,.
  \nonumber
\end{eqnarray}
According to Eq.~(\ref{de_7}) the symmetrical 
three-point function reads 
\begin{eqnarray} 
  \label{fin_3} 
  \gg^{(1)}_{\hat{A}_{\mu} \bar{q}_2 q_1}(p_2,p_1) 
  &=& 
  \g^{(1)}_{\hat{A}_{\mu} \bar{q}_2 q_1}(p_2,p_1) 
      + \g^{(1),CT}_{\hat{A}_{\mu} \bar{q}_2 q_1}(p_2,p_1)  
  \nonumber\\
  &=& 
  \g^{(1)}_{\hat{A}_{\mu} \bar{q}_2 q_1}(p_2,p_1) 
  -  T^0_{p_1,p_2} \g^{(1)}_{\hat{A}_{\mu} \bar{q}_2q_1}(p_2,p_1) 
  - \frac{e Q_q}{2}\left(
    \xi^{(1)}_{2,q_1} + \xi^{(1)}_{2,q_2}
  \right)
  \gamma^\mu
  \,, 
\end{eqnarray} 
and analogously for $\gg^{(1)}_{\bar{q}_2 q_1}$. 

Next to the fermionic three-point functions also the bosonic ones 
of~(\ref{gr:bsga}) have to be considered.
The corresponding WTIs read (see, e.g., \cite{amt_1}) 
\begin{eqnarray}
  \label{wti.2} 
  \lefteqn{\left. \frac{\delta^3 {\cal W}_{(\lambda)}(\g^{(1)})} 
  {\delta \lambda_A(-p_+ - p_-) \delta W^+_\rho(p_+)  \delta W^-_\sigma(p_-)} 
  \right|_{\phi=0} =} 
  \nonumber\\&&\mbox{} 
  i \left(p_{+} + p_{-}\right)^\mu \g^{(1)}_{\hat{A}_{\mu} W^{+}_{\rho} 
  W^{-}_{\sigma}} (p_{+},p_{-}) - i e \left(  \g^{(1)}_{W^{+}_{\rho} 
  W^{-}_{\sigma}} (p_{-}) - \g^{(1)}_{W^{+}_{\rho} W^{-}_{\sigma}} (-p_{+}) 
  \right) = 
  \Delta^{(1)}_{W,\hat{A}_{\mu} W^{+}_{\rho} W^{-}_{\sigma}}
  \,,\nonumber \\ 
  \label{wti.3} 
  \lefteqn{\left. \frac{\delta^3 {\cal W}_{(\lambda)}(\g^{(1)})} 
  {\delta \lambda_A(-p_+ - p_-) \delta G^+(p_+)  \delta G^-(p_-)} 
  \right|_{\phi=0} =} 
  \nonumber\\&&\mbox{} 
  i \left( p_{+} + p_{-} \right)^\mu \g^{(1)}_{\hat{A}_{\mu} G^{+} 
  G^{-}} (p_{+},p_{-}) - i e \left( \g^{(1)}_{G^{+} G^{-}} 
  (p_{-}) - \g^{(1)}_{G^{+} G^{-}}(-p_{+}) \right) 
  = \Delta^{(1)}_{W,\hat{A}_{\mu} G^{+} G^{-}}
  \,,\nonumber\\ 
  \label{wti.4} 
  \lefteqn{\left. \frac{\delta^3 {\cal W}_{(\lambda)}(\g^{(1)})} 
  {\delta \lambda_A(-p_+ - p_-) \delta W^+_\rho(p_+)  \delta G^-(p_-)} 
  \right|_{\phi=0} =} 
  \nonumber \\&&\mbox{} 
  i \left( p_{+} + p_{-} \right)^\mu \g^{(1)}_{\hat{A}_{\mu} 
  W^{+}_{\rho} G^{-} } (p_{+},p_{-})  - i e \left( \g^{(1)}_{ 
  W^{+}_{\rho} G^{-}} (p_{-}) - \g^{(1)}_{W^{+}_{\rho} G^{-}} (-p_{+}) 
  \right) = 
  \Delta^{(1)}_{W,\hat{A}_{\mu} W^{+}_{\rho} G^{-}}
  \,,
\end{eqnarray} 
where we have omitted their hermitian counterparts. 
Acting with 
$(1-T^2_{p_+,p_-})$ removes all the breaking terms on the r.h.s..
However, the two-point (quantum) Green functions in~(\ref{wti.4})
are not yet completely fixed as there is still freedom to add background 
gauge invariant counterterms
arising from STIs ($\Xi^S$) and normalization conditions ($\Xi^N$). 
In order to determine the counterterms $\Xi^S$ 
we first have to consider the WTIs for the two-point
Green functions with external background fields (see point~2)
as described in Section~\ref{sub:strategy}.
In a next step the Eqs.~(\ref{de_6}) are solved step-by-step and,
finally, the normalization conditions are implemented.

Note that due to the linearity of the WTIs the equations of this section
are not restricted to the one-loop order but have the same form 
at any order.

\vskip.3cm 
\noindent 
{\it 2. Background two-point functions}.
Before treating the quantum two-point functions of~(\ref{gr:bsgb})
we have to deal with the corresponding background counterparts.
They are constrained by the WTIs 
\begin{eqnarray}
  ip_\mu \g^{(1)}_{\hat{W}^+_\mu \hat{W}^-_\nu}(p) 
    + i M_W \g^{(1)}_{\hat{G}^+ \hat{W}^-_\nu}(p)
  \,\,=\,\, \Delta^{(1)}_{W,\lambda^+ \hat{W}^-_\mu}\,, 
  \nonumber \\ 
  ip_\mu \g^{(1)}_{\hat{W}^+_\mu \hat{G}^-}(p) 
  + i M_W \g^{(1)}_{\hat{G}^+\hat{G}^-}(p) 
  \,\,=\,\, \Delta^{(1)}_{W,\lambda^+ \hat{G}^-} 
  \,.
  \label{ep_8} 
\end{eqnarray} 
The r.h.s. of Eqs.(\ref{ep_8}) have mass dimension three 
as can be seen by power-counting arguments. 
Therefore we have to act from the left with the 
Taylor operator $(1-T^3_p)$ which leads to
\begin{eqnarray}\label{ep_9} 
  {\Psi}^{W,(1)}_{\lambda_+ \hat W_\mu^-} 
  &=& ip_\mu \gh^{(1)}_{\hat{W}^+_\mu \hat{W}^-_\nu}(p) 
    + i M_W \gh^{(1)}_{\hat{G}^+ \hat{W}^-_\nu}(p)
  \,\,=\,\, 
  iM_W
  \frac{1}{3!}
  p^\mu p^\rho p^\sigma 
  \partial_{p^\mu} \partial_{p^\rho} \partial_{p^\sigma}
  \g^{(1)}_{\hat{G}^+\hat{W}^-_\nu}(p)\Big|_{p=0} 
  \,, 
  \nonumber \\ 
  {\Psi}^{W,(1)}_{\lambda_+ \hat G^-} &=& 
  i p_\mu \gh^{(1)}_{\hat{W}^+_\mu\hat{G}^-}(p) 
  + i M_W \gh^{(1)}_{\hat{G}^+\hat{G}^-}(p) \,\,=\,\, 0 
  \,. 
\end{eqnarray} 
As a consequence, only the counterterm 
\begin{equation}
  \label{ep_9.1} 
  \Xi^{W,(1)}_{1} = 
  \int {\rm d}^4x  \left[ \xic{\partial W^2,1}^{W,(1)}\, \partial^\mu
  \hat{W}^{+}_{\mu} \partial^\nu  
  \hat{W}^{-}_{\nu} + \xic{\partial W^2,2}^{W,(1)}\, \partial^\mu
  \hat{W}^{+}_{\nu} \partial^\mu  
  \hat{W}^{-}_{\nu}  \right]
  \,,
\end{equation} 
is needed to restore the WTIs~(\ref{ep_8}). 
Notice that, in the case where the used  
renormalization scheme is invariant under background gauge 
symmetry, the breaking terms~(\ref{ep_9}) vanish and, 
therefore, only the STIs (which are discussed in point~3) have to
be restored. 

As in the previous section,
the same WTIs hold also here
to all loop orders, which is due to the linearity of
the equations.

\vskip.3cm 
\noindent 
{\it 3. Quantum two-point functions}. 
One of the main new features of the method presented in this paper
is the use of a particular set of STIs as derived in
Eqs.~(\ref{de_6}).
In the example presented at the end of Section~\ref{sub:strategy}
the last equation of~(\ref{de_6}) applied to the two-point
functions is shown in~(\ref{ex_co.1}).
In order to get the breaking terms on the r.h.s.
one has to proceed as indicated after~(\ref{de_6}).
In our case the differentiation of~(\ref{ST}) with respect to
$\Omega^\pm$ and $W^\mp$ or $G^\mp$
has to be performed. Altogether this leads to eight equations, which,
however, all have the same form. For
demonstration we only present one of them:
\begin{eqnarray} 
  \label{exap} 
  \lefteqn{\frac{\delta^{2} {\cal S}(\g)}{\delta \Omega^{+}_\nu(-p) \delta 
    W^{-}_{\mu}(p)}\bigg|_{\phi=0} =}
  \nonumber\\
  && \g_{\hat{W}^{+}_{\nu} {W}^{-}_{\mu}} - 
     \Big( g_{\nu\rho} + \g_{ W^{*,-}_{\rho} \Omega^{+}_\nu} \Big)  
       \g_{W^{+}_{\rho} W^{-}_{\mu}} 
     + \g_{\Omega^{+}_\nu  G^{*,-}}  \, \g_{G^{+} W^{-}_{\mu}} 
  \,\,=\,\, 
     \Delta^S_{\Omega^+_\nu W^-_\mu} 
  \,,
\end{eqnarray}
where the dependence on the external momenta is suppressed.
For our purposes only the one-loop approximation of these
equations are needed\footnote{Note that,  
  one can take into account that 
  $\g^{(0)}_{\hat{W}^{+}_{\nu} {W}^{-}_{\mu}}= 
   \g^{(0)}_{\hat{W}^{+}_{\nu} \hat{W}^{-}_{\mu}}= 
   \g^{(0)}_{{W}^{+}_{\nu} {W}^{-}_{\mu}}$ 
  and  analogously for $WG$ and $GG$ Green functions
  and that 
  the Green functions involving $\Omega$ and an anti-field
  vanish at tree level.}.
Furthermore, it is useful to combine two equations in such a
way that the two-point functions of a background and a quantum field
drop out.
After zero-momentum subtraction of the form $(1-T^2_p)$ one obtains 
\begin{eqnarray} 
  \label{exap.1}  
  \gh^{(1)}_{\hat{W}^{+}_{\nu} \hat{W}^{-}_{\mu}} 
  - \gh^{(1)}_{{W}^{+}_{\nu} {W}^{-}_{\mu}}
  \hphantom{xxxxxxxxxxxxxxxxxxxxxxxxxxxxxxxxxxxxxxxxxxxx}
  \nonumber\\\mbox{}
  -   \gh^{(1)}_{ W^{*,+}_{\rho} \Omega^{-}_\mu} 
      \g^{(0)}_{{W}^{+}_{\nu} {W}^{-}_{\rho}} 
  - 
      \gh^{(1)}_{ W^{*,-}_{\rho} \Omega^{+}_\nu} 
      \g^{(0)}_{{W}^{-}_{\mu} {W}^{+}_{\rho}} 
   -  \gh^{(1)}_{G^{*,+} \Omega^{-}_\mu} 
      \g^{(0)}_{{G}^{-} {W}^{+}_{\nu}} 
   -
           \gh^{(1)}_{G^{*,-} \Omega^{+}_\nu} 
           \g^{(0)}_{{G}^{+} {W}^{-}_{\mu}} 
  &=&\hat{\Psi}^{S,(1)}_{1,\nu\mu}
  \,, 
  \nonumber\\ 
  \gh^{(1)}_{\hat{W}^{+}_{\nu} \hat{G}^{-}} 
  - \gh^{(1)}_{{W}^{+}_{\nu} {G}^{-}}
  \hphantom{xxxxxxxxxxxxxxxxxxxxxxxxxxxxxxxxxxxxxxxxxxxxx}
  \nonumber\\\mbox{} 
  - \gh^{(1)}_{ W^{*,+}_{\rho} \Omega^{-}} 
  \g^{(0)}_{{W}^{+}_{\nu} {W}^{-}_{\rho}} 
  -\gh^{(1)}_{ G^{*,-} \Omega^{+}_\nu} 
  \g^{(0)}_{{G}^{+} {G}^{-}} 
  - \gh^{(1)}_{ W^{*,-}_{\rho} \Omega^{+}_\nu} 
    \g^{(0)}_{{G}^{-}{W}^{+}_\rho}
  -    
     \gh^{(1)}_{ G^{*,+} \Omega^{-}} 
     \g^{(0)}_{{G}^{-} {W}^{+}_\nu} 
  &=&\hat{\Psi}^{S,(1)}_{2,\nu} 
  \,, 
  \nonumber\\ 
  \gh^{(1)}_{\hat{G}^{+} \hat{G}^{-}} 
  - \gh^{(1)}_{{G}^{+} {G}^{-}}
  \hphantom{xxxxxxxxxxxxxxxxxxxxxxxxxxxxxxxxxxxxxxxxxxxxxx}
  \nonumber\\\mbox{} 
  -   \gh^{(1)}_{ W^{*,+}_{\rho} \Omega^{-}} 
      \g^{(0)}_{{G}^{+} {W}^{-}_{\rho}} 
  -     
      \gh^{(1)}_{ W^{*,-}_{\rho} \Omega^{+}} 
      \g^{(0)}_{{G}^{-} {W}^{+}_{\rho}} 
  -  \gh^{(1)}_{G^{*,+} \Omega^{-}} 
     \g^{(0)}_{{G}^{-} {G}^{+}} 
  -
          \gh^{(1)}_{G^{*,-} \Omega^{+}} 
          \g^{(0)}_{{G}^{+} {G}^{-}} 
  &=&\hat{\Psi}^{S,(1)}_{3}
  \,, 
  \nonumber \\ 
\end{eqnarray} 
where the equation containing 
$\gh^{(1)}_{{W}^{-}_{\nu} {G}^{+}}$
is not shown.
The breaking terms are given by 
\begin{eqnarray} 
  \Psi^{S,(1)}_{1,\nu\mu} &=& 
  - i \left(
   M^2_W p^\rho p^\sigma \partial_{p^\rho} \partial_{p^\sigma}
    \g^{(1)}_{ W^{*,-}_{\mu} \Omega^{+}_\nu}(p) + 
  2 M_W p_\mu p^\sigma \partial_{p^\sigma} \g^{(1)}_{ G^{*,-}
  \Omega^{+}_\nu}(p)
  \right)\Big|_{p=0}
  \,, 
  \nonumber\ \\  
  \Psi^{S,(1)}_{2,\nu} &=&
  - i M_W^2 p^\sigma \partial_{p^\sigma} 
  \g^{(1)}_{ W^{*,+}_\nu \Omega^{-}}(p)\Big|_{p=0}
  \,, 
  \nonumber\\ 
  \Psi^{S,(1)}_{3}  &=& 
  2 i M_W p^\nu p^\sigma \partial_{p^\sigma} 
  \g^{(1)}_{ W^{*,+}_\nu \Omega^{-}}(p)\Big|_{p=0}
  \,,
  \label{exap.2} 
\end{eqnarray} 
where the properties of the Green functions under
Hermitian conjugation have been used.  
Note again that the breaking terms in Eq.~(\ref{exap.2}) are finite and thus
do not depend on the regularization.

The breaking terms 
can be absorbed by counterterms 
for $\gh^{(1)}_{{W}^{+}_{\nu} {W}^{-}_{\mu}}$ and 
$\gh^{(1)}_{{G}^{+}{G}^{-}}$ of the form
\begin{eqnarray} 
  \label{exap.3} 
  \Xi^{S,(1)}_{1} 
  &=& \int {\rm d}^4x\Big[ \xic{\nabla V^2,1}^{S,(1)} \, \nabh^\mu W^{+}_{\mu} 
   \nabh^\nu W^{-}_{\nu} + \xic{\nabla V^2,2}^{S,(1)} \, \nabh^\mu W^{+}_{\nu} 
   \nabh^\mu W^{-}_{\nu}
  \nonumber \\&&\mbox{}
  +\xic{\nabla\Phi^2}^{S,(1)}  \, \nabh_\mu G^{+} 
  \nabh^\mu G^{-} + \left( \xic{\Phi^2,1} + \xic{\Phi^2\hat\Phi^2,1}
  v^2 \right)  G^+ G^-
  \nonumber \\&&\mbox{}
  +\xi^{S,(1)}_{\nabla \Phi V \Phi} \left( \hat\nabla^\mu  W^+_\mu G^- +
  \hat\nabla^\mu  W^-_\mu G^+ \right)
  \Big] 
  \,,
\end{eqnarray} 
which can be extracted from background gauge
invariant counterterms
of Eqs.~(\ref{cc_5}),~(\ref{cc_6}) and~(\ref{cc_7}) in
Appendix~\ref{app:cou}.

For the practical computation of the counterterm of Eq.~(\ref{exap.3})
we now use~(\ref{de_6}). Applied to two-point functions the third
equation can already be found in~(\ref{ex_co.1}). Adding the two
equations leads in our case to
\begin{eqnarray} 
  \label{new.exap.1} 
  \Xi^{S,(1)}_{1,\hat{W}^{+}_{\nu} \hat{W}^{-}_{\mu}} 
  - \Xi^{S,(1)}_{1, {W}^{+}_{\nu} {W}^{-}_{\mu}}
  -  \Big( \Xi^{S,(1)}_{ W^{*,+}_{\rho} \Omega^{-}_\mu}
  \g_{0,{W}^{+}_{\mu} {W}^{-}_{\rho}} +  
  \Xi^{S,(1)}_{ W^{*,-}_{\rho} \Omega^{+}_\nu} \g_{0,{W}^{-}_{\mu}
  {W}^{+}_{\rho}} \Big)  
  \nonumber\\\mbox{}
  - \Big(\Xi^{S,(1)}_{G^{*,+} \Omega^{-}_\mu} \g_{0,{G}^{-}
  {W}^{+}_{\nu}} +  
  \Xi^{S,(1)}_{G^{*,-} \Omega^{+}_\nu} \g_{0,{G}^{+}
  {W}^{-}_{\mu}} \Big)  
  &=&\Psi^{S,(1)}_{1,\nu\mu}
  \,.
\end{eqnarray} 
Similar equations are obtained for 
$\Xi^{S,(1)}_{1, {W}^{+}_{\nu} {G}^{-}}$
and
$\Xi^{S,(1)}_{1, {G}^{+} {G}^{-}}$
where $\Psi^{S,(1)}_{2,\nu}$ and $\Psi^{S,(1)}_{3}$ appear on
the r.h.s., respectively.

Before explicitly computing the coefficients $\xi^{S,(n)}_{i}$
in Eq.~(\ref{exap.3}), 
we have to determine the 
counterterms 
$\Xi^{S,(1)}_{ W^{*,+}_{\rho} \Omega^{-}_\mu}$ 
and 
$\Xi^{S,(1)}_{G^{*,+} \Omega^{-}_\mu}$ 
and perform
the renormalization 
of the Green functions 
$\gh^{(1)}_{W^{*,+}_{\rho}\Omega^{-}_\mu}$
and  
$\gh^{(1)}_{G^{*,+}\Omega^{-}_\mu}$.
This will be discussed in point~5 below. 
The renormalization of background Green functions in 
Eqs.~(\ref{exap.1}) is given by the WTIs~(\ref{ep_8}). 

Concerning the normalization conditions 
we are allowed to add the counterterm 
\begin{eqnarray} 
  \label{exap.4} 
    \Xi^{N,(1)}_{2} 
  &=& \int {\rm d}^4x
    \left[ -
    {\xi^{N,(1)}_{F^2} \over 4} \, 
          \Big( {F}^\alpha_{\mu\nu}(V + \hat V) 
          {F}^{\mu\nu}_\alpha (V + \hat V)\Big) 
  \right.\nonumber\\&&\left.\mbox{}
    +{\xi^{N,(1)}_{\nabla\Phi^2} \over 2} \, 
       \nabla_\mu (\Phi + \hat \Phi + v)^i 
       \nabla^\mu (\Phi + \hat \Phi + v)_i 
    \right] 
  \,.
  \nonumber 
\end{eqnarray} 
The coefficients $\xi^{N,(1)}_{F^2}$ and $\xi^{N,(1)}_{\nabla\Phi^2}$
are tuned  
in order to fix the mass of the $W$ boson, $M_W$, 
and the weak mixing angle $c_W$~\cite{msbkg}. 
Notice that the $\xi^{N,(1)}_{F^2}$ and $\xi^{N,(1)}_{\nabla\Phi^2}$ 
can be expressed as a combination of coefficients in
Eqs.~(\ref{cc_5}), (\ref{cc_6}),  
and (\ref{cc_7}), by requiring the BRST symmetry. 

\vskip.3cm 
This discussion completes the renormalization of the
background three-point functions. In the list of contributing 
Green functions, cf. Eqs.~(\ref{gr:bsga}) and~(\ref{gr:bsgb}),
only the quantum three-point functions and the ones involving ghosts
are missing. They will be treated in points~4 and~5.

\vskip.3cm 
\noindent 
{\it 4. Quantum three-point functions}. 
In this paragraph we consider the Green function of~(\ref{gr:bsga})
involving two fermions and a $W$ or a Goldstone boson.
According to our procedure we again have to consider the 
corresponding background Green functions first.

The background amplitude belonging to 
$\g^{(1)}_{{W}^{+}_{\nu} \bar{q} b}$
satisfies the one-loop identity of Eq.~(\ref{ex_1}).
The breaking and the counterterms are given by 
Eqs.~(\ref{ex_2}) and~(\ref{ex_3}), respectively.
Analogous equations hold for the 
Green function $\g^{(1)}_{{W}^{+}_{\nu} \bar{s} q}$
where $(b,q)$ is replaced by $(q,s)$.
Note that the counterterm $\Xi^{W,(1)}_{2}$ has no terms
involving a Goldstone-fermion vertex. Such contributions vanish through
the zero momentum subtraction.
The one-loop coefficients $\xic{\bar q b W,L/R}^{W,(1)} $ 
have been explicitly computed in~\cite{amt_2} in the case of QCD
corrections.
Note that the quark two-point functions have already been 
fixed by normalization conditions (cf. Eq.~(\ref{qua_nor})).

The STIs for the amplitudes $\g^{(1)}_{{W}^{+}_{\nu} \bar{q} b}$ 
and $\g^{(1)}_{{G}^{+} \bar{q} b}$
which correspond to the last equation of~(\ref{de_6})
are given by
\begin{eqnarray} 
  \label{eq:new_2}  
  \gh^{(1)}_{\hat{W}^{+}_{\nu} \bar{q} b}(p_q,p_b) - 
  \gh^{(1)}_{{W}^{+}_{\nu} \bar{q} b}(p_q,p_b) + 
  \gh^{(1)}_{\Omega^{+}_\nu W^{*,-}_{\rho}}(p_q+p_b)   
  \g^{(0)}_{{W}^{+}_{\rho} \bar{q} b} 
  \hphantom{xxxxxxxxxxxxxxxx}
  \nonumber\\\mbox{} 
  +\gh^{(1)}_{\Omega^{+}_\nu G^{*,-}}(p_q + p_b)
   \g^{(0)}_{\hat{G}^+\bar{q}b} 
  -\g^{(0)}_{\bar{q} q^\prime}(-p_q) 
   \g^{(1)}_{\Omega^{+}_\nu \bar{q}^{\prime*} b}( p_q,p_b) 
  -\g^{(1)}_{\Omega^{+}_\nu \bar{q} {q}^{\prime*}}(p_q,p_b) 
   \g^{(0)}_{\bar{q}^\prime b}(p_b) 
  &=& 
  \Psi^{S,(1)}_{\Omega^+_\nu \bar{q} b}
  \,, 
  \nonumber\\
  \gh^{(1)}_{\hat{G}^{+} \bar{q} b}(p_q,p_b) - 
  \gh^{(1)}_{{G}^{+} \bar{q} b}(p_q,p_b) + 
  \gh^{(1)}_{\Omega^{+} W^{*,-}_{\rho}}(p_q+p_b)   
  \g^{(0)}_{{W}^{+}_{\rho} \bar{q} b} 
  \hphantom{xxxxxxxxxxxxxxxxx}
  \nonumber\\\mbox{} 
  + \gh^{(1)}_{\Omega^{+}  G^{*,-}}(p_q+p_b)   
    \g^{(0)}_{\hat{G}^+\bar{q} b } 
  -\g^{(0)}_{ \bar{q} q^\prime}(-p_q) 
   \g^{(1)}_{\Omega^{+}  \bar{q}^{\prime*} b}( p_q,p_b) - 
   \g^{(1)}_{\Omega^{+}  \bar{q} {q}^{\prime*}}( p_q, p_b) 
   \g^{(0)}_{\bar{q}^\prime b}(p_b) 
  &=& 
  \Psi^{S,(1)}_{\Omega^+ \bar{q} b}  
  \,,
  \nonumber\\ 
\end{eqnarray} 
where zero momentum subtraction has already been applied.
They are obtained by considering the derivatives of 
Eq.~(\ref{ST}) with respect to $\bar{q}$, $b$ and 
$\Omega^{+}_\nu$ or $\Omega^{+}$.
The breaking terms are given by 
\begin{eqnarray} 
  \label{eq:new_3} 
  \Psi^{S,(1)}_{\Omega^+_\nu \bar{q} b} &=& 
  i \left(
  m_{q} \g^{(1)}_{\Omega^+_\nu \bar{q}^* b}(0,0) 
  + m_{b} \g^{(1)}_{\Omega^+_\nu\bar{q} b^*}(0,0)  
  \right)
  \,, 
  \nonumber \\ 
  \Psi^{S,(1)}_{\Omega^+ \bar{q} b} &=& 
  i \left(
  m_{q} \g^{(1)}_{\Omega^+ \bar{q}^* b}(0,0)  
  + m_{b} \g^{(1)}_{\Omega^+ \bar{q} b^*}(0,0)
  \right)
  \,, 
\end{eqnarray} 
where the Green functions on the r.h.s. are finite. 
They are removed by introducing a counterterm 
for the quantum fields (cf. Eq.~(\ref{cc_8}))
\begin{eqnarray} 
  \label{eq:new_4.1} 
  \Xi^{S,(1)}_{2} 
  &=& \int {\rm d}^4x 
  \left[
  \xic{\bar q b W,L }^{S,(1)} \, \bar q \not\!W^{+} P_L b   + 
  \xic{\bar q b W,R}^{S,(n)} \, \bar q \not\!W^{+} P_R \,b
  \right.\nonumber\\&&\left.\mbox{}
  +\xic{\bar q b G,L}^{S,(1)} \,  G^{+} \, \bar q P_L b + 
  \xic{\bar q b G,R}^{S,(n)} \,  G^{+} \, \bar q P_R \,b 
  + {\rm h.c.}\right] 
  \,, 
  \nonumber 
\end{eqnarray} 
where the values 
of the coefficients $\xi^{S,(1)}_{i}$
depend on the normalization of the Green functions 
involving an $\Omega$ field. 

We refrain from listing the equations that determine the
counterterms
$\Xi^{S,(1)}_{2,W^+_\mu\bar{q}b}$ 
and
$\Xi^{S,(1)}_{2,G^+\bar{q}b}$ 
(i.e. the ones corresponding to~(\ref{new.exap.1})) 
as the structure is similar to Eq.~(\ref{eq:new_2}).
The one-loop Green functions have to be replaced by the
corresponding counterterms. 

Note that the equations for the vertices involving the quarks
$s$ and $q$ are in complete analogy to the ones presented above.

\vskip.3cm
At this point of our analysis all the counterterms
for the quantum fields are expressed in terms of
the counterterms
$\Xi^{S,(1)}_{W^*\Omega}$
and
$\Xi^{S,(1)}_{G^*\Omega}$. They will be discussed below.

\vskip.3cm 
\noindent 
{\it 5. Ghost Green functions}. 
In the following we discuss the Green functions involving
Faddeev-Popov ghosts or the fields $\Omega$.

In the list of Green functions contributing to $b\to s\gamma$ at two
loops there are the amplitudes
$\g_{\hat A \bar c^\pm c^\mp}$
and
$\g_{\bar c^\pm c^\mp}$
which are related through the WTI
\begin{eqnarray} 
  \label{eq:new_5} 
  \lefteqn{\left. \frac{\delta^3 {\cal W}_{(\lambda)}(\g^{(1)})} 
  {\delta \lambda_A(-p_+ - p_-) \delta c^+(p_+)  \delta \bar c^{-}(p_-)} 
  \right|_{\phi=0} =} 
  \nonumber\\&&\mbox{} 
  i \left(p_{+} + p_{-}\right)^\mu \g^{(1)}_{\hat{A}_{\mu} c^{+} 
  \bar c^{-}} (p_{+},p_{-}) - i e \left(  \g^{(1)}_{c^{+} 
  \bar c^{-}} (p_{-}) - \g^{(1)}_{c^{+} \bar c^{-}} (-p_{+}) 
  \right) 
  \,\,=\,\, \Delta^{W}_{\lambda_A c^+ \bar c^-}
  \,, 
  \nonumber 
\end{eqnarray} 
and its Hermitian conjugate.
Acting with the Taylor operator $(1-T^1_{p_+,p_-})$ 
removes the breaking term $\Delta^{W}_{\lambda_A c^+ \bar c^-}$
and no counterterm is needed to restore the identity.
However, there is still freedom 
to add background gauge invariant counterterms to the amplitudes 
$\gh^{(1)}_{c^{\pm} \bar c^{\mp}}$.
The latter are related to those with external 
anti-fields by means of the Faddeev-Popov 
equations (cf. Appendix~\ref{app:ghost_equations}
and~\cite{grassi,krau_ew})
which read
\begin{eqnarray} 
  \label{e_2} 
  \g^{(1)}_{c_\pm \bar{c}_\mp}(p) \pm i p^{\mu}   
  \g^{(1)}_{c_\pm W^{*,\mp}_{\mu}}(p) - 
  \xi_W M_{\pm,G^{\mp}} \g^{(1)}_{c_\pm G^{*,\mp}}(p) 
  &=& 
  \Delta^{(1)}_{F, c^\pm \bar c^\mp}
  \,. 
\end{eqnarray} 
Thus the counterterms for 
$\g^{(1)}_{c_\pm \bar{c}_\mp}$ have to be chosen in such a way that
$\Delta^{(1)}_{F, c^\pm \bar c^\mp}$ is removed. 
In general they read
\begin{eqnarray} 
  \label{e_2.1} 
  \Xi^{F,(1)}_{1} &=& 
  \int {\rm d}^4 x \Big( \xi^{F,(1)}_1  
  \nabh_\mu  \bar c^+  \nabh_\mu  c^- + \xi^{F,(1)}_2 
  \bar c^+ c^-  + {\rm h.c.} \Big) 
  \,. 
\end{eqnarray} 
However, this does not completely fix the 
ghost two-point functions as also the anti-field-dependent 
Green functions in~(\ref{e_2}) have to be fixed.

In the remaining part of this subsection we discuss the 
renormalization of the missing two-point functions,
namely those involving $\Omega$ fields, like
$\gh^{(1)}_{\Omega^{\pm} G^{*,\mp}}$
and
$\gh^{(1)}_{\Omega^{\pm}_\mu W^{*,\mp}_{\nu}}$, 
and the ones with Faddeev-Popov ghosts and anti-fields, 
$\g^{(1)}_{c^\pm W^{*,\mp}_{\mu}}$
and
$\g^{(1)}_{c^\pm G^{*,\mp}}$. 
Actually, the following considerations 
are significantly simplified in the framework of 
dimensional regularization~\cite{hoo,maison}
(see~\cite{martin} for a practical calculation),
since only those identities involving fermions
will produce breaking terms and since
there is no tree-level coupling of fermions with ghost fields.
However, we present the general analysis
as outlined in Section~\ref{sub:strategy}. 
Also at higher orders this part of the discussion will be useful 
even in dimensional regularization.

In a first step we want to mention that the Green functions 
$\g^{(1)}_{c^{\mp} W^{*,\pm}_{\mu}}$, 
$\g^{(1)}_{c^{A} W^{*,3}_{\mu}}$,
$\g^{(1)}_{c^{Z} W^{*,3}_{\mu}}$ and
$\g^{(1)}_{c^{*,3} c^+ c^-}$
are fixed by normalization conditions for the wave function of the
ghost fields. A convenient choice corresponds to
\begin{eqnarray} 
  \label{gh_nor} 
  -i \left. \partial_{p^\mu}  \gg_{c^{A} W^{*,3}_{\mu}} 
  \right|_{p=0} \,\,=\,\, s_W
  \,,
  &~~~~~& 
  \left. \gg_{c^{*,3} c^+ c^-}\right|_{p^+=p^-=0} = -ie\frac{c_W}{s_W}
  \,. 
\end{eqnarray} 
Then the WTI 
\begin{eqnarray} 
  &&  i (p+q)_\nu \g^{(1)}_{\hat W^+_\nu W^{*,3}_\mu c^-}(p,q) + 
  i M_W \g^{(1)}_{\hat G^+ W^{*,3}_\mu c^-}(p,q)  
  \,\,=\,\, 
  \nonumber\\ 
  &&\hspace{.2cm} 
  + i \frac{e}{s_W} \left( \g^{(1)}_{ W^{*,+}_{\mu} c^-}(q) +
  s_W \g^{(1)}_{ W^{3,*}_{\mu} c^A}(-p) - 
  c_W \g^{(1)}_{ W^{3,*}_{\mu} c^Z}(-p) \right)  +
  \Delta^{W,(1)}_{\lambda_+ W^{*,3}_\mu c^-}
  \,,
  \label{gh_1} 
\end{eqnarray} 
can be used to obtain the counterterm for
$\g^{(1)}_{\hat W^+_\nu W^{*,3}_\mu c^-}$
and for $\g^{(1)}_{W^{*,\mp}_\mu c^\pm }$
which removes the breaking term $\Delta^{W,(1)}_{\lambda_+ W^{*,3}_\mu c^-}$.
On the other hand, from the STI 
\begin{eqnarray} 
  \label{gh_2} 
  &&  
    i (p+q)_\nu \g^{(1)}_{W^+_\nu W^{*,3}_\mu c^-}(p,q) 
  + i q_\nu \g^{(1)}_{W^-_\nu W^{*,3}_\mu c^+}(p,-p-q) 
  \nonumber\\&&\mbox{}
  + i M_W \g^{(1)}_{G^+ W^{*,3}_\mu c^-}(p,q) 
  + i M_W \g^{(1)}_{G^- W^{*,3}_\mu c^+}(p,-p-q) 
  \,\,=\,\, 
  \nonumber \\ 
  &&\hspace{1.5cm}
  \mbox{}
  + i\frac{e}{s_W} 
  \left(  \g^{(1)}_{ W^{*,+}_{\mu} c^-}(q) 
        + \g^{(1)}_{ W^{*,-}_{\mu} c^+}(-p-q) 
        + s_W \g^{(1)}_{ W^{3,*}_{\mu} c^A}(-p) 
        - c_W \g^{(1)}_{ W^{3,*}_{\mu} c^Z}(-p) 
  \right)
  \nonumber \\ 
  &&\hspace{1.5cm}\mbox{}
  - i p_\mu \g^{(1)}_{c^{*,3} c^+ c^-}(-p-q,q) 
  + \Delta^{S,(1)}_{ c^+ W^{*,3}_\mu c^-}
  \,, 
\end{eqnarray}
the Green function $\g^{(1)}_{W^+_\nu W^{*,3}_\mu c^-}$ is determined
and the counterterms given in Eqs.~(\ref{cc_3}) remove the breaking term 
$\Delta^{S,(1)}_{ W^{*,3}_\mu c^+ c^-}$.
Finally, in the STI
\begin{eqnarray}
  && 
  \g^{(1)}_{\hat W^+_\nu W^{*,3}_\mu c^-}(p,q) - 
  \g^{(1)}_{W^+_\nu W^{*,3}_\mu c^-}(p,q) + 
  i\frac{e}{s_W}  \g^{(1)}_{\Omega^{+}_\nu W^{*,-}_{\mu}}(p)
  \,\,=\,\, 
  \nonumber \\ 
  &&\mbox{} \hspace{2cm} 
  - i q^\rho \g^{(1)}_{W^-_\rho W^{*,3}_\mu \Omega^+_\nu}(p,-p-q) - 
  iM_W \g^{(1)}_{G^- W^{*,3}_\mu \Omega^+_\nu}(p,-p-q) 
  \nonumber \\ 
  &&\mbox{} \hspace{2cm} 
  -i p_\mu \g^{(1)}_{c^{*,3} \Omega^+_\nu c^-}(-p-q,q)
  + \Delta^{S,(1)}_{\Omega^+_\nu W^{*,3}_\mu c^-}
  \,,
  \label{gh_22} 
\end{eqnarray} 
we only have finite Green functions on the r.h.s. and thus
from this equation it is possible to determine
the renormalization of $\g^{(1)}_{\Omega^{+}_\nu W^{*,-}_{\mu}}$. 

In principle, also for
$\g^{(1)}_{\Omega^{+}_\nu W^{*,-}_{\mu}}$
a normalization condition can be chosen.
However, the corresponding parameters in Eq.~(\ref{cc_4})
are automatically fixed by the symmetries of the theory.

A complete equivalent system can be derived to treat
the two-point functions with a scalar $\Omega$ field and with
$W^{*,-}_\mu$ replaced by $G^{*,-}$.
In the latter case
the renormalization of  
$\g^{(1)}_{c^{-} G^{*,+}}$ and $\g^{(1)}_{c^{i} G^{*,0}}$ ($i=A,Z$)
is needed in order to fix the ghost mass parameters.
For a detailed discussion we refer to~\cite{grassi,krau_ew}.

The equations presented in this paragraph follow the method outlined
in Section~\ref{sub:strategy}.
In particular, Eq.~(\ref{gh_1}) fixes the Green function
$\g^{(1)}_{\hat W^+_\nu W^{*,3}_\mu c^-}(p,q)$
which involves the background field $\hat W^+_\nu$.
Furthermore,
Eqs.~(\ref{gh_2}) and~(\ref{gh_22}), which correspond to the first and
second equations of~(\ref{de_6}), respectively, contain
the corresponding quantum Green functions and 
Eqs.~(\ref{gh_nor}) fix the normalization conditions.

Notice that the renormalization of 
the amplitudes with external anti-fields and 
ghost fields is fairly arbitrary for the present computation.
These normalization conditions do not influence
the physical observables  
in the process $b\rightarrow s \gamma$.

\vskip.3cm 
\noindent 
{\it 6. QED gauge coupling}. 
Besides the normalization conditions discussed above, 
we have to treat the QED coupling constant as remaining free parameter.
Its renormalization is achieved by introducing the counterterm 
\begin{eqnarray} 
  \label{QED} 
  \Xi^{N,(1)}_{3} 
  &=& \int {\rm d}^4x \left[- {\xic{F^2,3}^{N,(1)} \over 4} \, 
  {F}_{\mu\nu}(V + \hat V) {F}^{\mu\nu}(V + \hat V) \right] 
  \,,
  \nonumber 
\end{eqnarray} 
where ${F}_{\mu\nu}(V + \hat V) $ is the abelian field strength. 
 
\vskip.7cm

As already mentioned above the analysis presented in this section is
quite general. Thus, in order to conclude we 
briefly mention the simplifications due to
the use of Dimensional Regularization
accompanied with the 't~Hooft-Veltman definition of
$\gamma_5$~\cite{hoo,maison}. 

First, all identities which do not involve fermion lines are
preserved. In particular the breaking terms of Eqs.~(\ref{exap.1})
are zero as only diagrams involving virtual ghost particles contribute.
For the same reason the ghost sector is highly simplified as there is
no fermionic contribution to 
Eqs.~(\ref{gh_1}),~(\ref{gh_2}) and~(\ref{gh_22}).
Furthermore, those breaking terms to be computed from diagrams 
which don't involve a chiral vertex are also zero.  In the above
analysis this would correspond to Eq.~(\ref{bsg1}).

As a final remark, we would like to emphasize that owing a
background gauge invariant regularization, 
Eqs.~(\ref{bsg1}), (\ref{wti.4}), (\ref{ep_8}), (\ref{ex_1}), 
(\ref{e_2}) and~(\ref{gh_1}) 
are automatically preserved
and only the Eqs.~(\ref{exap.1}), (\ref{eq:new_2}), 
(\ref{gh_2}) and~(\ref{gh_22}) have to be studied in detail.

From the analysis performed in the example of this section, 
it is clear that the strategy outlined in Section~\ref{sub:strategy} 
can effectively be applied to each
process of the SM. 
One can also see, that 
the zero-momentum subtraction significantly simplifies the practical
computation of the non-invariant counterterms.  

 
\section{\label{sec:con}Conclusions} 

In this paper a general procedure to perform the renormalization using
the algebraic
renormalization and the background field method is discussed. It is
shown that the computational
problems to evaluate the non-invariant counterterms within a generic
subtraction scheme can be
drastically reduced in the framework of the BFM and by means of an
intermediate subtraction.

Recently, several progress have been done in constructing new
regularization schemes for chiral gauge theories at 
perturbative~\cite{luscher,pernici,Jeg00}
and at non-perturbative level~\cite{Gol00}. However, at the
practical
level, the algebraic renormalization with a non-invariant regularization
scheme in our formulation turns out to be still superiour.
In addition, owing a scheme which is explicitly background gauge
invariant only few counterterms (cf. Appendix~\ref{app:cou}) 
are indeed necessary to restore the STIs.
Finally, even in case that a scheme would exist which is invariant under
all the symmetries of the SM, the present paper provides a complete
analysis of the relations between Green function with external
background fields and those with quantum ones.

To summarize, we want to give a brief outline for possible 
applications of the method. In general,  
the method can be divided into the following two main steps: 

\begin{enumerate} 
\item {\it Anti-field-dependent 
  counterterms $\Xi\ton_{\#}[\phi,\hat\phi,\phi^*,\Omega]$.} 

In a first step one considers the WTIs and fixes all  
possible counterterms using the 
gauge symmetry constraints. This amounts 
to determining  the coefficients of the 
counterterms 
$\Xi^{W,(n)}_\#[\phi,\hat\phi,\phi^*,\Omega]$ 
(see, for instance, Eq.~(\ref{gh_1})). 

The second step concerns the computation of the counterterms 
which are left after fixing the WTIs, 
namely $\Xi^{S,(n)}_\#[\phi,\hat\phi,\phi^*,\Omega]$ 
(see, e.g., Eqs.~(\ref{gh_2}) and~(\ref{gh_22})). 
In particular, to restore the relation between the background and 
the quantum fields 
(cf. the second and the third equations of~(\ref{de_6}) and
in the example Eq.~(\ref{gh_22}))
one needs to study STIs 
obtained by differentiating Eq.~(\ref{ST}) with respect to one 
anti-field and two ghost fields and  with respect to one 
anti-field, $\Omega$ and one ghost field 
(cf. the first and the second equations of~(\ref{de_6}) 
and Eq.~(\ref{gh_22})). Finally, the remaining free parameters 
$\Xi^{N,(n)}_\#[\phi,\hat\phi,\phi^*,\Omega]$ 
are fixed by normalization conditions (cf. Eqs.~(\ref{gh_nor})). 
 
\item {\it Anti-field-independent counterterms $\Xi\ton_{O}[\phi,\hat\phi]$.} 
 
The anti-field-independent counterterms constitute the main part of the 
diagram computations.
At first, one fixes all the possible counterterms using the 
background gauge symmetry. This is done by exploiting the 
corresponding WTI and computing  $\Xi^{W,(n)}_O[\phi,\hat\phi]$ 
(compare Eqs.~(\ref{ex_1}),~(\ref{bsg1}),~(\ref{wti.3}) and~(\ref{ep_8})). 
Then, one restores the STIs by 
$\Xi^{S}_{O}[\phi,\hat\phi]$ (see Eqs.~(\ref{exap}),~(\ref{exap.1}), 
and ~(\ref{eq:new_2})), and finally 
the free parameters $\Xi^{N}_{O}[\phi,\hat\phi]$ 
can be tuned on the physical data by means of normalization conditions. 
 
Note that, as is shown in the example, in a first step 
the two-point functions are considered.
Afterwards the 
three-point functions are fixed, then the four-point functions etc..
The success of this procedure is guaranteed by the
consistency conditions between ${\cal S}_0$  
and ${\cal W}_{(\lambda)}$ as outlined in Appendix~C of \cite{amt_1}. 
\end{enumerate}

 
\section*{Acknowledgements} 
We thank Carlo Becchi and Paolo Gambino
for comments and suggestions. 
The research of P.A.G. is under the NSF grants nos. PHY-9722083 and 
PHY-0070787. 
 
 
\renewcommand {\theequation}{\Alph{section}.\arabic{equation}} 
\begin{appendix} 
 
 
\setcounter{equation}{0} 
\section{Linearized Slavnov-Taylor operator, coupling of $\Omega$
 fields and 
 functional Taylor operator} 
\label{app:linear} 
 
The linearized Slavnov-Taylor operator for a generic functional 
${\cal F}$ is given by 
\begin{eqnarray} 
{\cal S}_{\gg} (\cf) & \equiv &  \int {\rm d}^4 x \left\{ 
\left( s_W \partial_\mu c_Z + c_W \partial_\mu c_A \right) 
\left( s_W \frac{\delta \cf}{\delta Z_\mu} + 
c_W \frac{\delta \cf}{\delta A_\mu} \right) \right. 
\nonumber \\&&\mbox{} 
+ \sum_{\alp=A,Z,\pm,a} b_{\alp}  \frac{\delta \cf}{\delta \bar{c}^{\alp}} 
+ (\gg, \cf) + (\cf,\gg)
 \nonumber\\&&\mbox{} 
+ \Omega^{3}_\mu 
  \left[ c_W  \left(  \frac{\delta{\cal F}}{\delta \hat{Z}_\mu} -
\frac{\delta{\cal F}}{\delta Z_\mu}\right)  
 -  s_W \left( \frac{\delta{\cal F}}{\delta \hat{A}_\mu}      -
\frac{\delta{\cal F}}{\delta A_\mu} \right) \right]  
\nonumber\\&&\mbox{} 
+\Omega^{\pm}_\mu \left(  \frac{\delta{\cal F}}{\delta \hat{W}^\pm_\mu} - 
\frac{\delta{\cal F}}{\delta W^\pm_\mu}\right) 
+\Omega ^a_\mu  \left(  \frac{\delta{\cal F}}{\delta \hat{G}^a_\mu} - 
\frac{\delta{\cal F}}{\delta G^a_\mu}\right) 
\nonumber\\&&\left. \mbox{} 
+\Omega^{\pm} \left( \frac{\delta{\cal F}}{\delta \hat{G}^\pm} -
\frac{\delta{\cal F}}{\delta G^\pm}\right)  
+\Omega^0  \left( \frac{\delta{\cal F}}{\delta \hat{G}^0} -
\frac{\delta{\cal F}}{\delta {G}^0}\right)  
+\Omega^H  \left( \frac{\delta{\cal F}}{\delta \hat{H}} -
\frac{\delta{\cal F}}{\delta H}\right) \right\}  
\,, 
\label{line_ST} 
\end{eqnarray} 
where 
\begin{eqnarray}\label{bracket} 
\left( X, Y \right) &=& \int {\rm d}^4x \, \left[ 
\frac{\delta X}{\delta W^{*,3}_\mu}  \frac{\delta Y}{\delta W^3_\mu} + 
\frac{\delta X}{\delta W^{*,\pm}_\mu}  \frac{\delta Y}{\delta W^\mp_\mu} + 
\frac{\delta X}{\delta G^{*,a}_\mu}  \frac{\delta Y}{\delta G^a_\mu} + 
\frac{\delta X}{\delta c^{*,\pm}}  \frac{\delta Y}{\delta c^\mp} + 
\frac{\delta X}{\delta c^{*,3}} \frac{\delta Y}{\delta c^3} 
\nonumber\right.\\&&\left.\mbox{} 
+ 
\frac{\delta X}{\delta c^{*,a}}  \frac{\delta Y}{\delta c^a} + 
\frac{\delta X}{\delta c^{*,\pm}}  \frac{\delta Y}{\delta c^\mp} + 
\frac{\delta X}{\delta G^{*,\pm}}  \frac{\delta Y}{\delta G^\mp} + 
\frac{\delta X}{\delta G^{*,0}}  \frac{\delta Y}{\delta G^0} + 
\frac{\delta X}{\delta H^*}  \frac{\delta Y}{\delta H} 
\nonumber\right.\\&&\left.\mbox{} 
+ 
\sum_{I=L,Q,u,d,e} \left( 
\frac{\delta X}{\delta \bar{\psi}^{*I}} \frac{\delta Y}{\delta \psi^{I}} + 
\frac{\delta X}{\delta {\psi}^{*I}} \frac{\delta Y}{\delta \bar{\psi}^{I}}
\right) \right] 
\,. 
\end{eqnarray} 
Since ${\cal S}(\gg) =0$, the 
operator ${\cal S}_{\gg}$ is nilpotent. We also introduce the tree-level 
linearized operator ${\cal S}_0\equiv{\cal S}_{\g_0}$ 
where $\g_0$ is the tree-level action. 
 
In the method discussed in this paper
the couplings of the $\Omega$-fields 
(${\cal L}_{\Phi\Pi, \Omega}$) represent an important 
piece of information. In fact they are essential in the derivation the STIs 
for the counterterms involving background fields. In Ref.~\cite{amt_1}, the 
general building blocks of the Lagrangian with 
anti-fields has been given and the 
couplings with  $\Omega$-fields are contained in Eqs.~(A.2) and~(A.3) 
of Appendix A of~\cite{amt_1}. 
However, for the convenience of the reader, we present the 
couplings ${\cal L}_{\Phi\Pi, \Omega}$ in the explicit form: 
\begin{eqnarray} 
  \label{cou_Om} 
  {\cal L}_{\Phi\Pi, \Omega} 
    &=& 
  {\Omega}^{3}_{\mu} \left\{ \left( {c}_W \partial_{\mu} \bar{c}^{Z} -   s_W 
      \partial_{\mu} \bar{c}^{A} \right) - 
    {i\, e \over s_{W}} \left[ \left( 
        W^{+}_{\mu} + \hat{W}^{+}_{\mu} \right) \bar{c}^{-} - \left( 
        W^{-}_{\mu} + \hat{W}^{-}_{\mu} \right) \bar{c}^{+} \right] \right\} 
  \nonumber \\&&\mbox{} 
  + 
  \Omega^{\mp}_{\mu}  \left\{\!\partial_{\mu} \bar{c}^{\pm} \mp i e \left( 
      W^{\pm}_{\mu} + \hat{W}^{\pm}_{\mu} \right) \left( \bar{c}^{A} - 
      \frac{c_{W}}{s_{W}} \bar{c}^{Z} \right) 
  \right.\nonumber\\&&\left.\qquad\mbox{} 
  \pm i e \bar{c}^{\pm}\left[ 
      \left( A_{\mu} + \hat{A}_{\mu} \right) - \!\frac{c_{W}}{s_{W}} 
      \left( Z_{\mu} + \hat{Z}_{\mu} \right) \right] \right\} 
  \nonumber \\&&\mbox{} 
  + 
  \Omega^{a}_{\mu} \left\{ \partial_{\mu} \bar{c}^{a} - g_{s} f^{abc} 
    \left( G^{b}_{\mu} + \hat{G}^{b}_{\mu} \right) \bar{c}^{c} \right\} 
 \nonumber \\&&\mbox{} 
  +   \Omega^H \left\{  \frac{i e \, \xi_W}{2 s_{W}} \left[ \left(G^{+} + 
        \hat{G}^{+}\right) \bar{c}^{-} - 
      \left(G^{-} + \hat{G}^{-}\right) \bar{c}^{+} \right] + 
    \frac{e \, \xi_Z}{2 s_{W} c_{W}} 
    \left( G^{0} + \hat{G}^{0} \right) \bar{c}^{Z} \right\} 
  \nonumber \\&&\mbox{} 
  +   \Omega^{\mp} \left\{ \pm \frac{i e \, \xi_W }{2 s_{W}} 
    \left[ H + \hat{H} + v \pm i 
      \left( G^{0} + \hat{G}^{0} \right) \right] \bar{c}^{\pm} 
  \right.\nonumber\\&&\left.\qquad\mbox{} 
    \mp i e \left(G^{\pm} + 
      \hat{G}^{\pm}\right)\left(\xi_A \bar{c}^{A} - 
      \xi_Z\, \frac{c^{2}_{W} - s^{2}_{W}}{2 c_{W} s_{W}} \bar{c}^{Z} 
    \right) \right\} 
  \nonumber \\&&\mbox{} 
  + \Omega^{0} \left\{  \frac{e \, \xi_W}{2 s_{W}} \left[ 
      \left(G^{+} + 
        \hat{G}^{+}\right) \bar{c}^{-} + 
      \left(G^{-} + \hat{G}^{-}\right) \bar{c}^{+} \right] - 
    \frac{e\, \xi_Z}{2 s_{W} c_{W}} 
    \left( H + \hat{H} + v \right) \bar{c}^{Z} \right\} \,,
  \nonumber\\ 
\end{eqnarray} 
where $\bar{c}^Z,\bar{c}^A,\bar{c}^\pm$ and $\bar{c}^a$ 
are the anti-ghost fields. 
Notice that the Feynman rules for these new vertices are related to the 
couplings of the anti-fields with 
quantum fields (cf. Eq.~(A.1) of~\cite{amt_1}) 
simply by exchanging the ghost fields with the anti-ghost fields. 

The Taylor operator $T^\delta$ of the functional $\Gamma$ is defined as 
follows. 
One first considers the relevant amplitude which results from functional 
derivatives with respect to fields denoted by subscripts 
$\Gamma_{\phi_1(p_1)\phi_2(p_2)\ldots\phi_m(p_m)}$ 
with 
$\sum_{j=1}^{m}p_j=0$. 
Then  the Taylor expansion $T^{\delta}$ 
in the independent momenta up to degree $\delta$ acts formally 
as 
\begin{eqnarray} 
  T^\delta \Gamma &=& \sum_{m=1}^{\infty} \int \prod_{i=1}^{m} {\rm d}^4 p_i 
  \phi_i (p_i) \delta^4 ({\scriptstyle \sum_{j=1}^{m}} p_j) 
  T^{\delta}_{p_1,\dots,p_m} 
  \left . 
  \Gamma_{\phi_1(p_1)\phi_2(p_2)...\phi_m(p_m)} 
  \right |_{\sum_{j=1}^{m}p_j=0} 
  \label{definitionT2} 
  \,. 
\end{eqnarray} 
A remarkable property of $T^\delta$ is that $T^{\delta_1} T^{\delta_2} = 
T^{\delta}$ with $\delta={\rm min}\{\delta_1,\delta_2\}$. 
Note that the Taylor operator is scale-invariant, but it does not commute 
with spontaneous symmetry breaking. 

 
\setcounter{equation}{0} 
\section{Auxiliary functional constraints} 
\label{app:ghost_equations} 
 
We recall that the SM in the BFM~\cite{grassi} is 
completely  defined in terms of the following functional identities (up to 
normalization conditions) 
\begin{enumerate} 
\item The Nakanishi-Lautrup identities (\ref{eq:gau_fix}), which implement the 
  gauge fixing conditions to all orders, 
\item the Abelian Anti-ghost Equation 
  (in the case of BFM see second reference in~\cite{grassi}, Eq.~(4.28)), 
\item the non-abelian WTI given in Eq.~(\ref{WT}) 
  for the background gauge invariance,
\item the STI given by Eq.~(\ref{ST}) for the BRST symmetry, 
\item the Faddeev-Popov equations of motion (see \cite{grassi,krau_ew}), and 
\item the abelian WTI given by Eq.~(\ref{WT}) for the background 
  gauge invariance restricted to the $U(1)$ factor. 
\end{enumerate} 
 
The sets 1 to 4 of functional identities are 
imposed on the theory by requiring 
invariance under the corresponding symmetries. 
The sets 5 and 6 are derived constraints 
of the commutation relations of previous functional identities. However, 
for practical purposes, they turn out to be relevant. 
In the following, we briefly  discuss 
this issue. For more details we refer to the literature. 
 
For a generic non-invariant scheme all
possible functional identities can be 
spoiled by local breaking terms and the 
method presented in this paper can be applied 
to all of them. In addition, since all the identities are linear 
(except the STI), the intermediate subtraction at 
zero momentum 
drastically reduces the number of breaking terms and the latter
can easily be removed by breaking terms. 

In this paper
we emphasized the role of the WTIs and the STIs, however, also the 
supplementary constraints should be taken into account. 
As a consequence, 
it is possible
to define a reduced functional generator and to renormalize certain 
anti-field-dependent amplitude in terms of 
Green functions with external ghost fields. 
 
The Nakanishi-Lautrup identities~(\ref{eq:gau_fix}) are discussed in detail 
in~\cite{grassi,krau_ew}. 
In~\cite{krau_ew}, the gauge fixing depends only on the 
scalar background fields $\hat\Phi$. 
There a complete discussion for the renormalization 
of  (\ref{eq:gau_fix}) is presented. In~\cite{grassi}, 
the same analysis is performed in the 
background `t Hooft  gauge fixing. In particular, we notice that the zero 
momentum subtraction already removes 
all the possible breaking terms and no additional 
non-invariant counterterms are indeed needed. 
 
In the case of the SM, the STIs and the gauge fixing conditions are not 
sufficient to fix 
the abelian sector of the theory~\cite{grassi,krau_ew} completely.
In addition, one has to use another functional equation 
which controls the renormalization of the abelian ghost fields. 
Including also the background fields the equation reads
\begin{eqnarray}
  \label{AAE_phy} 
  \lefteqn{
  c_W \frac{\delta \g}{\delta c_A} + s_W \frac{\delta \g}{\delta c_Z} + 
  \frac{i e}{2 c_W} \left( \hat{G}^+ \frac{\delta \g}{\delta\Omega^+} - 
  \hat{G}^- \frac{ \delta \g}{\delta\Omega^-} \right) - 
  \frac{e}{2 c_W} \left( \hat{G}_0 \frac{\delta \g}{\delta\Omega^H} - 
  (\hat{H}+v) \frac{\delta \g}{\delta\Omega^0} \right) =} 
  \nonumber \\ 
  && 
  {e \over 2 c_W} \left( H^* G_0 - 
  G^*_0 (H+v)\right)  
  + {i e \over 2 c_W} \left(G^{+,*} G^- -  G^{-,*} G^+ \right)
  \nonumber\\ 
  &&\mbox{}  + 
  \sum_{\alpha} \left( \frac{1}{6} \bar{Q}^{L,*}_{\alp} Q^{L}_{\alpha} + 
  {2 \over 3} \bar{u}^{R,*}_{\alpha} u^{R}_{\alpha} -  {1 \over 3} 
  \bar{d}^{R,*}_{\alpha} d^{R}_{\alpha} - {1 \over 2} \bar{L}^{L,*}_{\alpha} 
  L^{L}_{\alpha} - \bar{e}^{R,*}_{\alpha} e^{R}_{\alp} \right)  +{\rm h.c.} 
  \nonumber\\ 
  &&\mbox{} + \left(s_W \partial^2 \bar c^{Z} 
  + c_W  \partial^2 \bar c^{A} \right)
  \,,  
\end{eqnarray} 
where we used the notation of~\cite{amt_1}. 
Notice that Eq.~(\ref{AAE_phy}) 
is linear in $\g$. However, by means of zero momentum 
subtraction this equation is spoiled. This is due to the fact that the 
UV power counting of the ghost fields $c_A$ and $c_Z$ and the one of $\Omega$ 
are different and, consequently, over-subtractions are generated. 
Fortunately, the algebraic analysis of this problem is simple.

The Faddeev-Popov equations of motion should be 
analyzed along the same lines. 
The complete analysis has been given 
in~\cite{grassi,krau_ew,fg}. In the application 
discussed in Section~\ref{sub:application} of the present paper, 
a particular set of Faddeev-Popov 
equations of motion has been used (cf. Eq.~(\ref{e_2})) 
and their renormalization was analyzed. 

The Lagrange multiplier equations (\ref{eq:gau_fix}), the Faddeev-Popov
equations of motion and
the Abelian Anti-ghost equation (\ref{AAE_phy}) are linear differential
functional equations
which can be solved by simple redefinitions of the functional
$\g_0$ and of the
anti-fields. This is because the set of the functional differential
operators associated with those supplementary 
constraints\footnote{E.g., the functional operator  
${\delta }/{\delta b_C}$ acting on $\g$ is associated with 
Eq.~(\ref{eq:gau_fix}).}
forms a sub-algebra of the complete algebra including the WTIs and
STIs.
The solution of the renormalized supplementary equations is called
the {\it reduced functional}~\cite{brs,zinn,itzy}. At tree level it
reads
\begin{eqnarray}
  \label{eq:red_func}
  \g^{\rm red.func.}_0 &=& \g_0 - \int {\rm d}^4x 
  \left[ b_C {\cal F}^{C}(V,\Phi,b)
  + \bar c\, \partial^2\, c \,  \right. 
  \nonumber \\&&
  \left. \mbox{}
  + \Omega^{\alpha}_\mu \left( \nabla^\mu  \bar c \right)_\alpha +
  \Omega^{a}_\mu \left( \nabla^\mu  \bar c \right)_a +
  \Omega^i \bar c_\alpha t^\alpha_{ij} \left( \Phi + \hat \Phi + v \right)^j
  \right. \nonumber \\&& \left.\mbox{}
  + \Phi^{*,i} c \, t^0_{ij} \left(  \Phi + \hat \Phi + v \right)^j
  + \left(\bar \psi^{*,I}  c \, T^0_{IJ}  \psi^J + {\rm h.c.}\right)
  \right] 
  \,,
\end{eqnarray}
where $C$, $\alpha$ and $a$ are the indices 
for the adjoint representation of $SU(3) \times SU(2) \times U(1)$,
$SU(2)$ and $SU(3)$, respectively. The ghost $c$ belongs
to the $U(1)$ sector. It can
also be written in terms of the rotated fields $c = c_W c_A + s_W c_Z$. The
reduced functional $ \g^{\rm red.func.}_0 $
does not depend on $b_C$, 
on the anti-ghosts $\bar c^{C}$, 
on the abelian ghost $c$
or on the $\Omega$ fields. In Eq.~(\ref{eq:red_func}), we used the compact
notation introduced in Section~\ref{sec:sti} and used in
Appendix~\ref{app:cou}.
The explicit form of the second line can be found in
Eq.~(\ref{cou_Om}) of Appendix~\ref{app:linear} and the explicit form of the
third line can be read off from Eq.~(A.1) of Ref.~\cite{amt_1} 
by selecting the contributions of
the abelian ghost.
$\g^{\rm red.func.}_0$ depends only on the following combinations
\begin{eqnarray}
  \label{eq:an_red_func}
  \tilde V^{*,C}_\mu &=& V^{*,C}_\mu + \left( \hat \nabla_\mu \bar c
  \right)^C\,, 
  \nonumber  \\
  \tilde \Phi^{*,i} &=&  \tilde \Phi^{*,i} + \bar c_\alpha t^{\alpha i}_{~~j}
  \left( \hat \Phi + v \right)^j  + \bar c \, t^{0 i}_{~~j} \left( \hat \Phi +
  v \right)^j
  \,,
\end{eqnarray}
of the anti-fields $V^{*,C}_\mu$ and $\Phi^{*,i}$ where $ t^0_{ij} $ is the
$U(1)$ generator in the
representation of the scalar fields.
At higher orders, some suitable normalization conditions
should be taken into 
account in order to avoid
spurious off-shell IR problems~\cite{grassi,krau_ew}.
Note, that the superscript ``red.func.'' is omitted in the main text
of the paper.
 
 
\setcounter{equation}{0} 
\section{Background gauge invariant counterterms for the STIs} 
\label{app:cou} 

In the present appendix, we list and classify all possible 
counterterms $\Xi^S$ needed to restore the STIs.  We assume that the WTIs 
for the background gauge invariance are already recovered. 
Furthermore, we assume that all other constraints such as the 
Nakanishi-Lautrup identities, 
the Faddeev-Popov equations and the anti-ghost equation 
(cf. Appendix~\ref{app:ghost_equations}) are 
satisfied. This implies that we can consider the 
simple factors of the gauge group separately from the abelian one, and 
the dependence upon anti-ghost, Lagrangians multiplier and abelian ghost 
field is already taken into account.
 
In the description of the general counterterms, we follow the previous 
classification  into anti-field dependent and independent counterterms, 
namely   $\Xi^S_{\#}[\phi,\hat\phi,\phi^*,\Omega]$          and 
$\Xi^S_{O}[\phi,\hat\phi]$, respectively.  
In addition, we organize the anti-field 
dependent counterterms according to the highest ghost number. 

We use the following
notation for the counterterms
\begin{eqnarray}
  \label{nota}\nonumber
  \int {\rm d}^4x \, 
  \xic{\phi^1 \phi^2 \dots \phi^n, i} \, 
  T^{a_1 a_2 \dots a_n}_i \,\, 
  \phi_{a_1}^1 \phi_{a_2}^2 \dots \phi_{a_n}^n\,,
\end{eqnarray}
where $\phi_{a_i}^i$
denotes the fields and their derivatives and
$T^{a_1 a_2 \dots a_n}_i$ 
contains  field-independent Lorentz and gauge group invariant 
(in the adjoint or in the matter representation) tensors whose 
independent component are parameterized by the 
index $i$. $\xic{\phi^1 \phi^2 \dots \phi^n, i}$ are the 
coefficients of the counterterms. Eventually, 
$\xic{\phi^1 \phi^2 \dots \phi^n, i}^{(n)}$ denotes
the $n^{\rm th}$ contribution to the 
coefficient $\xic{\phi^1 \phi^2 \dots \phi^n, i}$. 
In Section~\ref{sub:application} also 
the notation 
$\xic{\phi^1 \phi^2 \dots \phi^n, i}^{W,(n)}$, 
$\xic{\phi^1 \phi^2 \dots \phi^n, i}^{S,(n)}$ 
and $\xic{\phi^1 \phi^2 \dots \phi^n, i}^{N,(n)}$ 
is introduced to distinguish between the counterterms arising from  
the WTIs, the STIs and the normalization conditions, respectively.

\subsubsection*{Anti-field dependent terms   
  $\Xi^S_{\#}[\phi,\hat\phi,\phi^*,\Omega]$} 
 
\begin{enumerate} 
\item {\it Anti-ghost fields} 

The most negative ghost number is carried by the anti-fields of the 
ghost fields, $c^*_3$ and $c^*_\pm$, therefore the counterterms of the type 
\begin{eqnarray} 
  \label{cc_1} 
 \int {\rm d}^4x 
 \left[ 
 \xic{c^* c^2,1} \, \epsilon_{\alpha \beta \gamma}  \, 
 c^{*,\alpha} c^\beta c^\gamma 
  + \xic{c^* c^2,2} \, f_{abc} \, c^{*,a} c^b c^c 
  \right] 
 \,,
\end{eqnarray} 
are the most general background gauge invariant contribution. 
Note that due to the Abelian Anti-ghost equation~(\ref{AAE_phy}), 
the dependence on the abelian ghost field has already been fixed. 
Here $\epsilon_{\alpha \beta \gamma}$ and 
$f_{abc}$ are the structure constants  of the $su(2)$ 
and $su(3)$ algebras, respectively. 

Notice that the parameters $\xi_{c^*c^2,i}$ $(i=1,2)$ correspond to
the wave function renormalization of the $SU(2)$ and $SU(3)$ ghost
fields, respectively.
Therefore
they are fixed by normalization conditions of the type~(\ref{gh_nor}).
 
\item{\it Couplings of anti-fields with ghost fields and background fields} 
 
Having fixed the anti-fields of the ghosts, we now turn
to the anti-fields of the quantum fields. 
Therefore one has to select the couplings of 
the anti-fields for the gauge fields, $W^{*,3}_\mu$ and $W^{*,\pm}_\mu$,
and of the anti-fields of the scalar fields, $H^*, G^*_0$ and $G^*_\pm$, 
with the ghost fields. 
A generic counterterm can be expressed by the equation 
\begin{eqnarray} 
  \label{cc_2} 
  \int {\rm d}^4x 
  \left[ 
  \xic{V^{*} \!c,1}  \, 
    V^{*,\alpha \mu} \left(\hat\nabla_\mu c\right)_\alpha + 
  \xic{V^{*} \!c,2}  \, 
    V^{*,a \mu} \left( \hat\nabla_\mu c\right)_a + 
  \xic{\Phi^{*} \!c} \, 
    \Phi^{*,i} c_\alpha t^{\alpha}_{ij} (\hat \Phi + v)^j 
  \right] 
  \,, 
\end{eqnarray} 
where $t^{\alpha}_{ij}$ are $SU(2)$ generators in the scalar 
representation.  Notice again that, due to Eq.~(\ref{AAE_phy}), the 
coupling of the scalar fields with the abelian ghost field are already 
determined. Therefore, in Eq.~(\ref{cc_2}) only the coupling 
between $SU(2)$-ghost fields and scalars has been taken into account. The 
covariant derivatives are defined by 
$\left( \hat\nabla_\mu c\right)^a = 
  \partial_\mu c^a - f^{a}_{~bc} \hat V^b_\mu c^c$ and 
$\left(\hat\nabla_\mu c\right)^\alpha = 
  \partial_\mu c^\alpha -\epsilon^{\alpha}_{~\beta \gamma} 
  \hat V^\beta_\mu c^\gamma$. 

The parameters $\xi_{V^*c,i}$ $(i=1,2)$ and $\xi_{\Phi^*c}$ amount to a
wave function
renormalization of the $SU(2)$ and $SU(3)$ quantum gauge fields and of the
quantum scalar multiplet
$\Phi^i$. Consequently, they are fixed by normalization conditions 
like~(\ref{gh_nor})
instead of STIs. This simplifies further the task of the computation.
 
\item{\it Couplings of anti-fields with ghost fields and quantum fields} 
 
The gauge fields $V^\alpha_\mu$ and 
$V^a_\mu$ transform as vectors of the adjoint representation
under background gauge transformations. Thus, among the 
anti-field dependent counterterms, 
we also have to list the following contributions 
\begin{eqnarray} 
  \label{cc_3} 
&&\int {\rm d}^4x 
\Big[ 
  \xic{ V^{*} \!V c,1} \epsilon_{\alpha \beta \gamma} \, 
    V^{*,\alpha \mu} V^\beta_\mu c^\gamma + 
  \xic{V^{*} \!V c,2} \, f_{a b c}  V^{*,a \mu} V^b_\mu c^c 
  \nonumber \\ && \hspace{0.8cm} 
  \mbox{}+
  \xic{V^{*} \!V c,3} d_{a b c}  \, V^{*,a \mu} V^b_\mu c^c + 
  \xic{\Phi^{*} \!c} \Phi^{*,i} c_\alpha t^{\alpha}_{ij} \Phi^j 
  \nonumber \\ && \hspace{0.8cm} 
  \mbox{}+
  \sum_\psi\Big(
  \xic{\bar\psi^* \psi,1}\bar\psi^*_I T_{\alpha}^{IJ}\psi_J c^\alpha + 
  \xic{\bar\psi \psi^*,1} \bar\psi_{I} 
  T_{\alpha}^{IJ,\dagger}\psi^*_J c^\alpha 
  \nonumber \\ && \hspace{0.8cm}  
  \mbox{}+
  \xic{\bar\psi^* \psi,2}\bar\psi^*_I T_{a}^{IJ}\psi_J c^a + 
  \xic{\bar\psi \psi^*,2} \bar\psi_{I} T_{a}^{IJ,\dagger}\psi^*_J c^a 
  + \mbox{h.c.}\Big)
  \Big]\,, 
  \end{eqnarray} 
where $T_{\alpha}^{IJ}= T_{\alpha}^{L,IJ} P_L + 
  T_{\alpha}^{R,IJ} P_R$ and $T_{a}^{IJ}$ are the generators for $SU(2)$ 
and $SU(3)$ gauge transformations, respectively. Notice that 
$T^{\dagger}$ is the 
pseudo-hermitian conjugate of $T$.
In order to take into account the 
mixings among fermion generations in
the counterterm of the type   
$\xic{\bar\psi^* \psi,2}\bar\psi^*_I T_{a}^{IJ}\psi_J c^a$ 
a summation is understood. For instance,
\begin{eqnarray}
  \label{summ_qu_1}
  \sum_\psi \int {\rm d}^4x  
  \, \xic{\bar\psi^* \psi,2}\bar\psi^*_I T_{a}^{IJ}\psi_J c^a &=& 
  \sum_{q,q'=u,c,t,d,s,b} \int {\rm d}^4x   
  \, \xic{\bar q^* q',2} \, \bar q^* \frac{i \, \lambda_a}{2} q' \, c^a\,,
\end{eqnarray}
where $\lambda_a$ are the $SU(3)$ Gell-Mann matrices and 
$\xic{\bar q^* q',1}$ is a complex matrix. 

\item {\it Couplings of anti-fields with $\Omega$} 
 
A well-known feature of the BFM is the fact that 
multiplets are renormalized by the same constant. 
As a consequence, 
the renormalization of the background  
and of the quantum fields is related to each other. The counterterms that 
control these relations at the quantum levels are 
\begin{eqnarray} 
  \label{cc_4} 
  \int {\rm d}^4x 
  \left[ 
  \xic{V^* \Omega,1} V^{*, \alpha \mu} \Omega_{\alpha \mu} + 
  \xic{V^* \Omega,2 } V^{*,a \mu} \Omega_{a \mu} + 
  \xic{\Phi^* \Omega} \Phi^{*,i} \Omega_i 
  \right]\,. 
\end{eqnarray} 

\end{enumerate} 
 
\subsubsection*{Anti-field independent terms $\Xi^S_{O}[\phi,\hat\phi]$} 
 
\begin{enumerate} 
\item {\it Gauge sector} 
 
We have to recall that abelian quantum gauge field $V_\mu$ has no 
background partner. More precisely, combining the WTI for the 
abelian factor and the anti-ghost equation (cf. Eq.~(\ref{AAE_phy}) 
in Appendix~\ref{app:ghost_equations}), the 
couplings with background abelian gauge fields 
$\hat V_\mu$ are completely fixed~\cite{grassi,krau_ew}. 
In the following formulae, only the background fields 
$\hat V^\alpha_\mu$ and  $\hat V^a_\mu$, 
for the $SU(2)$ and $SU(3)$ part of the gauge group, 
are taken into account.  
The most general counterterm containing only gauge fields reads 
\begin{eqnarray} 
  \label{cc_5} 
  && 
  \int {\rm d}^4x 
  \Big[ 
\xic{F^2,1} \,\hat F^{\alpha,\mu\nu} \hat F_{\alpha,\mu\nu} +
\xic{F^2,2} \,\hat F^{a,\mu\nu} \hat F_{a,\mu\nu} +
\xic{F^2,3} \, F^{\mu\nu}  F_{\mu\nu} 
\nonumber \\
&&\hspace{0.8cm}
\mbox{}+
\xic{FVV,1} \, \epsilon_{\alpha \beta \gamma} \hat F^{\alpha,\mu\nu}
V^\beta_{\mu} V^\gamma_{\nu} +
\xic{FVV,2}  \, f_{a b c}  \hat F^{a,\mu\nu}  V^b_{\mu} V^c_{\nu} 
\nonumber \\
&&\hspace{0.8cm}
\mbox{}+
\xic{F\nabla V,1} \,\hat F^{\alpha,\mu\nu} \left(\hat\nabla_{\mu} V_{\nu}
\right)_{\alpha} +
\xic{F\nabla V,2} \,\hat F^{a,\mu\nu} \left(\hat\nabla_{\mu} V_{\nu}
\right)_{a} 
\nonumber \\
&&\hspace{0.8cm}
\mbox{}+
\xic{\nabla V^2,1} \left( \hat\nabla_\mu V_{\nu} \right)^{\alpha}
\left(\hat\nabla^\mu V^{\nu}\right)_{\alpha} +
\xic{\nabla V^2,2} \left(\hat\nabla^\mu V_{\mu}\right)^{\alpha}
\left(\hat\nabla^\nu V_{\nu}\right)_{\alpha} 
\nonumber \\
&&\hspace{0.8cm}
\mbox{}+
\xic{\epsilon FVV,1}
\epsilon_{\alpha \beta \gamma}
\epsilon^{\mu\nu\rho\sigma} \hat F_{\alpha,\mu\nu}V^b_{\beta}
V^\gamma_{\sigma} +
\xic{\epsilon FVV,2}
f_{a b c}
\epsilon^{\mu\nu\rho\sigma} \hat F_{a,\mu\nu}V^b_{b} V^c_{\sigma}
\nonumber \\
&&\hspace{0.8cm}
\mbox{}+
\xic{\nabla V^2,3} \left(\hat\nabla_\mu V_{\nu}\right)^{a}
\left(\hat\nabla^\mu V^{\nu}\right)_{a} +
\xic{\nabla V^2,4} \left(\hat\nabla^\mu V_{\mu}\right)^{a}
\left(\hat\nabla^\nu V_{\nu}\right)_a 
\nonumber \\
&&\hspace{0.8cm}
\mbox{}+
\xic{V^2\nabla V,1}
\, \epsilon_{\alpha \beta \gamma}  \left(\hat\nabla^\mu
V^{\nu}\right)^{\alpha} V^\beta_{\mu} V^\gamma_{\nu} +
\xic{V^2\nabla V,2}
\, f_{a b c} \left(\hat\nabla^\mu V^{\nu}\right)^{a}  V^b_{\mu} V^c_{\nu} 
\nonumber \\
&&\hspace{0.8cm}
\mbox{}+
\xic{V^2\nabla V,3} \, d_{a b c} \left(\hat\nabla^\mu V^{\nu}\right)^{a}
V^b_{\mu} V^c_{\nu} +
\xic{V^2\nabla V,4} \, d_{a b c} \left(\hat\nabla^\mu V_{\mu}\right)^{a}
V^b_{\nu} V^{c,\nu} 
\nonumber \\
&&\hspace{0.8cm}
\mbox{}+
\xic{V^4,1} \,  V^{\beta,\mu} V^{\gamma,\nu}  V_{\beta,\mu} V_{\gamma,\nu} +
\xic{V^4,2} \,  V^{\beta,\mu} V^{\gamma,\nu}  V_{\beta,\nu} V_{\gamma,\mu} 
\nonumber \\
&&\hspace{0.8cm}
\mbox{}+
\xic{V^4,3}  \,  V^{b,\mu} V^{a,\nu}  V_{b,\mu} V_{a,\nu} +
\xic{V^4,4}  \,  V^{b,\mu} V^{a,\nu}  V_{b,\nu} V_{a,\mu} 
\nonumber \\
&&\hspace{0.8cm}
\mbox{}+
\xic{V^4,5} \, d_{ab}^{~x} d_{x cd} V^{b,\mu} V^{a,\nu} V_{c,\mu} V_{d,\nu} 
\nonumber \\
&&\hspace{0.8cm}
\mbox{}+
\xic{V^4,6} \, V^{a,\mu} V^{a,\nu} V_{\alpha,\mu} V_{\alpha,\nu} +
\xic{V^4,7} \, V^{a,\mu} V^{a,\mu} V_{\alpha,\nu} V_{\alpha,\nu} 
\nonumber \\
&&\hspace{0.8cm}
\mbox{}+
\xic{V^4,8} \, f_{ab}^{~x} f_{x cd} V^{b,\mu} V^{a,\nu} V_{c,\mu} V_{d,\nu} +
\xic{V^4,9} \, \epsilon_{\alpha \beta}^{~x} 
\epsilon_{x \gamma \delta} V^{\beta,\mu}
V^{\alpha,\nu} V_{\gamma,\mu} V_{\delta,\nu} 
\nonumber \\
&&\hspace{0.8cm}
\mbox{}+
\xic{V^2,1} \,  V^{\beta,\mu} V_{\beta,\mu} +
\xic{V^2,2} \,  V^{b,\mu}  V_{b,\mu}
  \Big]\,, 
\end{eqnarray} 
where $\hat F^{\alpha,\mu\nu} $ and $\hat F^{a,\mu\nu} $ 
are the $SU(2)$ and $SU(3)$ background gauge field strengths, respectively. 
$F^{\mu\nu}$ is the 
abelian quantum gauge field strength and
$d^{abc}$ is the totally symmetric tensor in the 
adjoint representation of $su(3)$. 
 
It is clear that at the one-loop order 
only the quantum field independent counterterms 
contribute. This means that only coefficients $\xic{F^2,1},
\xic{F^2,1}$ and  
$\xic{F^2,3}$ in the first line of Eq.~(\ref{cc_5}) are needed. At
two loops, the inspection of  
diagrams reveals that also the counterterms quadratic in the quantum 
fields, namely $\xic{F V^2,1}, \dots, \xic{\nabla V^2,4},\xic{V^2,1}$
and $\xic{V^2,1}$, are necessary.  
At higher orders, all the other coefficients become important. 
 
\item {\it Mixed scalar and gauge sector} 
 
Due to the absence of the background partner
to the abelian gauge field $V_\mu$,  
the covariant derivative of scalars $\Phi_i$ and fermions $\psi_I$ is
defined with respect to  
the abelian quantum 
gauge field and the $SU(2)\times SU(3)$ background gauge fields: 
$\hat\nabla_\mu \Phi_i = \partial_\mu  \Phi_i - V_\mu t^0_{ij} \Phi^j
- \hat V_{\alpha\mu} t^\alpha_{ij} \Phi^j$ and  
$\hat\nabla_\mu \psi_I = \partial_\mu  \psi_I - V_\mu T^0_{IJ} \psi^J
- \hat V_{\alpha\mu} T^\alpha_{IJ} \psi^J  
- \hat V_{a\mu} T^a_{IJ} \psi^J$. Here $t^0_{ij}$ and  $T^0_{IJ}$ are
the hypercharge generators  
in the scalar and fermion representations. 
 
The general counterterm for the kinetic terms for scalars and
their interaction with gauge fields  
is described by the following expression 
\begin{eqnarray} 
  \label{cc_6} 
  &&
  \int {\rm d}^4x 
  \Big[ 
  \xic{\nabla\hat\Phi \nabla\hat\Phi}  \left(\hat\nabla^\mu (\hat\Phi
  + v) \right)^{i} \left(\hat\nabla_\mu (\hat\Phi + v) \right)_{i} +  
  \xic{\nabla\hat\Phi \nabla\Phi}   \left(\hat\nabla^\mu (\hat\Phi + v)
  \right)^{i} \left(\hat\nabla_\mu \Phi  \right)_{i}
  \nonumber \\ &&\hspace{0.8cm} 
  \mbox{}+
  \xic{\nabla\hat\Phi V \hat\Phi} \left(\hat\nabla^\mu (\hat\Phi + v)
  \right)^{i} t^\alpha_{ij}  V_{\alpha,\mu} \left(\hat\Phi + v
  \right)^j +  
  \xic{\nabla\hat\Phi V \Phi}  \left(\hat\nabla^\mu (\hat\Phi + v)
  \right)^{i} t^\alpha_{ij}  V_{\alpha,\mu} \Phi^j   
  \nonumber \\ &&\hspace{0.8cm} 
  \mbox{}+
  \xic{\nabla\Phi \nabla\Phi}  \left(\hat\nabla^\mu \Phi\right)^{i}
  \left(\hat\nabla_\mu \Phi\right)_{i} +  
  \xic{\nabla\Phi V \hat\Phi} \left(\hat\nabla^\mu \Phi\right)^{i}
  t^\alpha_{ij}  V_{\alpha,\mu} \left(\hat\Phi + v \right)^j 
  \nonumber \\ &&\hspace{0.8cm} 
  \mbox{}+
  \xic{\nabla\Phi V \Phi}  \left(\hat\nabla^\mu \Phi\right)^{i}
  t^\alpha_{ij}  V_{\alpha,\mu} \Phi^j +  
  \xic{V^2 \Phi\hat\Phi,1} \, t^\alpha_{ij}  V_{\alpha,\mu} \Phi^j
  t^{\beta,i}_{~k}  V_{\beta}^{\mu}\left(\hat\Phi + v \right)^k 
  \nonumber \\ &&\hspace{0.8cm} 
  \mbox{}+
  \xic{V^2 \hat\Phi^2,1}   \, t^\alpha_{ij}  V_{\alpha,\mu}
  \left(\hat\Phi + v\right)^{j}  t^{\beta,i}_{~k}
  V_{\beta}^{\mu}\left(\hat\Phi + v \right)^k +  
  \xic{V^2 \Phi^2,1} 
  \, t^\alpha_{ij}  V_{\alpha,\mu} \Phi^j t^{\beta,i}_{~k}
  V_{\beta}^{\mu} \Phi^k  
  \nonumber \\ &&\hspace{0.8cm} 
  \mbox{}+
  \xic{V^2 \hat\Phi^2,2} \,  V_{\alpha,\mu}   V^{\alpha,\mu}
  \left(\hat\Phi + v\right)_{j} \left(\hat\Phi + v \right)^j +  
  \xic{V^2 \Phi\hat\Phi,2} \,  V_{\alpha,\mu}   V^{\alpha,\mu}
  \Phi_{j} \left(\hat\Phi + v \right)^j 
  \nonumber \\ &&\hspace{0.8cm} 
  \mbox{}+
  \xic{V^2 \Phi^2,2}  \,  V_{\alpha,\mu}   V^{\alpha,\mu} \Phi_{j} \Phi^j + 
  \xic{V^2 \hat\Phi^2,3} \,  V_{a,\mu}   V^{a,\mu} \, \left(\hat\Phi +
  v\right)_{j} \left(\hat\Phi + v \right)^j 
  \nonumber \\ &&\hspace{0.8cm} 
  \mbox{}+
  \xic{V^2 \Phi\hat\Phi,3}  \,  V_{a,\mu}   V^{a,\mu} \, \Phi_{j}
  \left(\hat\Phi + v \right)^j +  
  \xic{V^2 \Phi^2,3} \,  V_{a,\mu}   V^{a,\mu} \, \Phi_{j} \Phi^j 
  \Big]
  \,.
  \end{eqnarray} 
Note that the last three terms are a 
consequence of the background gauge invariance in the $SU(3)$ part of
the gauge group.  
 
\item {\it Scalar sector} 
 
To complete the counterterms for the scalar sector, we must list
the ones that reconstruct the correct scalar potential 
\begin{eqnarray} 
  \label{cc_7} 
  &&
  \int {\rm d}^4x 
  \Big[
  \xic{\Phi^2}   \,  \Phi_{j} \Phi^j + 
   \xic{\hat\Phi^2} \,  \left(\hat\Phi + v\right)_{j} \left(\hat\Phi +
  v \right)^j +  
  \xic{\Phi\hat\Phi} \,  \Phi_{j} \left(\hat\Phi + v \right)^j
  \nonumber \\ &&\hspace{0.8cm} 
  \mbox{}+
  \xic{\Phi^4} \,  \left( \Phi_{j} \Phi^j \right)^2 + 
  \xic{\hat\Phi^4} \,  \left[\left(\hat\Phi + v\right)_{j}
  \left(\hat\Phi + v \right)^j \right]^2  +  
  \xic{\Phi\hat\Phi^2,1} \,  \left[\Phi_{j} \left(\hat\Phi + v
  \right)^j  \right]^2  
  \nonumber \\ &&\hspace{0.8cm} 
  \mbox{}+
  \xic{\Phi\hat\Phi^2,2} \,   \Phi_{j} \Phi^j  \left(\hat\Phi +
  v\right)_{k} \left(\hat\Phi + v \right)^k +  
  \xic{\Phi^3\hat\Phi}  \,   \Phi_{j} \left(\hat\Phi + v \right)^j
  \Phi_k \Phi^k
  \nonumber \\ &&\hspace{0.8cm}
  \mbox{}+
  \xic{\hat\Phi^3\Phi}   \,   \Phi_{j} \left(\hat\Phi + v \right)^j
  \left(\hat\Phi + v\right)_k \left(\hat\Phi + v \right)^k  
  \Big]
  \,. 
\end{eqnarray} 
Note that at actually one- and two-loop order only few of these terms
are needed.
 
\item {\it Fermion sector} 
 
Finally, we have to discuss the fermionic terms. Again, it is easy to
establish the most general background gauge invariant contributions 
\begin{eqnarray} 
  \label{cc_8} 
  &&
  \sum_{\psi \psi'}
  \int {\rm d}^4 x 
  \Big[ 
  \xic{\bar\psi \nabla \psi^\prime} \bar\psi_I \hat\nabla^{IJ}
  \psi_J^\prime 
  + 
  \xic{\bar\psi\psi^\prime V,1} \bar\psi_I T_{\alpha}^{IJ}  \not\! V^\alpha
  \psi_J^\prime +  
  \xic{\bar\psi\psi^\prime V, 2} \bar\psi_I T_{a}^{IJ}  \not\! V^a
  \psi_J^\prime   
  \nonumber \\ && \mbox{} 
  +
  \xic{\bar\psi \psi^\prime \hat\Phi,m} Y^{IJ,j}_m \bar\psi_I \psi_J^\prime
  \left(\hat\Phi + v \right)_j +  
  \xic{\bar\psi \psi^\prime \Phi,m} Y^{IJ,j}_m \bar\psi_I
  \psi_J^\prime 
  \Phi_j 
  +\mbox{h.c.}
  \Big]\,. 
\end{eqnarray} 
Here, the fermionic indices run over the $SU(2)$ isospin, the color
and the flavours of fermions. In order to simplify the notation, we
introduced $Y^{IJ,j}_m$  
to denote $m$ independent couplings between scalars and fermions. These 
tensors satisfy the relations
$T_{\alpha,K}^{~I} Y^{KJ,j}_m + Y^{IK,j}_m T_{\alpha,K}^{~J} +
t_{\alpha,r}^{~i} Y^{IJ,r}_m = 0$,
$T_{0,K}^{~I} Y^{KJ,j}_m + Y^{IK,j}_m T_{0,K}^{~J} + t_{0,r}^{~i}
Y^{IJ,r}_m = 0$
and
$T_{a,K}^{~I} Y^{KJ,j}_m + Y^{IK,j}_m T_{a,K}^{~J} = 0$ for every $m$.  

\end{enumerate} 

\end{appendix} 

 
\def\ap#1#2#3{{\it Ann. Phys. (NY)} {\bf #1} (#2) #3} 
\def\jmp#1#2#3{{\it  J. Math. Phys.} {\bf #1} (#2) #3} 
\def\rmp#1#2#3{{\it Rev. Mod. Phys.} {\bf #1} (#2) #3} 
 
\def\app#1#2#3{{\it Act.~Phys.~Pol.~}{\bf B #1} (#2) #3} 
\def\apa#1#2#3{{\it Act.~Phys.~Austr.~}{\bf#1} (#2) #3} 
\def\cmp#1#2#3{{\it Comm.~Math. Phys.~}{\bf #1} (#2) #3} 
\def\cpc#1#2#3{{\it Comp.~Phys.~Commun.~}{\bf #1} (#2) #3} 
\def\epjc#1#2#3{{\it Eur.\ Phys.\ J.\ }{\bf C #1} (#2) #3} 
\def\fortp#1#2#3{{\it Fortschr.~Phys.~}{\bf#1} (#2) #3} 
\def\ijmpc#1#2#3{{\it Int.~J.~Mod.~Phys.~}{\bf C #1} (#2) #3} 
\def\ijmpa#1#2#3{{\it Int.~J.~Mod.~Phys.~}{\bf A #1} (#2) #3} 
\def\jcp#1#2#3{{\it J.~Comp.~Phys.~}{\bf #1} (#2) #3} 
\def\jetp#1#2#3{{\it JETP~Lett.~}{\bf #1} (#2) #3} 
\def\mpl#1#2#3{{\it Mod.~Phys.~Lett.~}{\bf A #1} (#2) #3} 
\def\nima#1#2#3{{\it Nucl.~Inst.~Meth.~}{\bf A #1} (#2) #3} 
\def\npb#1#2#3{{\it Nucl.~Phys.~}{\bf B #1} (#2) #3} 
\def\nca#1#2#3{{\it Nuovo~Cim.~}{\bf #1A} (#2) #3} 
\def\plb#1#2#3{{\it Phys.~Lett.~}{\bf B #1} (#2) #3} 
\def\prc#1#2#3{{\it Phys.~Reports }{\bf #1} (#2) #3} 
\def\prd#1#2#3{{\it Phys.~Rev.~}{\bf D #1} (#2) #3} 
\def\pR#1#2#3{{\it Phys.~Rev.~}{\bf #1} (#2) #3} 
\def\prl#1#2#3{{\it Phys.~Rev.~Lett.~}{\bf #1} (#2) #3} 
\def\pr#1#2#3{{\it Phys.~Reports }{\bf #1} (#2) #3} 
\def\ptp#1#2#3{{\it Prog.~Theor.~Phys.~}{\bf #1} (#2) #3} 
\def\sovnp#1#2#3{{\it Sov.~J. Nucl. Phys.~}{\bf #1} (#2) #3} 
\def\tmf#1#2#3{{\it Teor.~Mat.~Fiz.~}{\bf #1} (#2) #3} 
\def\yadfiz#1#2#3{{\it Yad.~Fiz.~}{\bf #1} (#2) #3} 
\def\zpc#1#2#3{{\it Z.~Phys.~}{\bf C #1} (#2) #3} 
\def\ppnp#1#2#3{{\it Prog.~Part.~Nucl.~Phys.~}{\bf #1} (#2) #3} 
\def\ibid#1#2#3{{ibid.~}{\bf #1} (#2) #3} 
 
%
%
 
\end{document}